\documentclass[5p,times]{elsarticle}

\usepackage{lineno}
\usepackage{hyperref}

\usepackage{adjustbox}
\usepackage{diagbox}
\usepackage{booktabs,graphicx}
\usepackage{listings}

\usepackage{color}
\usepackage{xcolor}
\usepackage{algorithm}
\usepackage{algpseudocode}
\usepackage{amsmath}
\usepackage{amssymb}
\usepackage{color, colortbl}
\usepackage{hhline}

\usepackage{tikz}
\usepackage{xcolor}
\usetikzlibrary{arrows.meta}
\definecolor{vat-blue}{RGB}{128,184,224}
\definecolor{vat-yellow}{RGB}{255,224,128}
\definecolor{vat-green}{RGB}{128,216,168}
\definecolor{vat-purple}{RGB}{219,113,235}
\definecolor{vat-grey}{RGB}{231,229,231}
\definecolor{vat-common-green}{RGB}{145,220,90}
\definecolor{vat-optional-blue}{RGB}{37,183,211}
\definecolor{vat-alternative-yellow}{RGB}{255,210,77}
\definecolor{vat-alternative-blue}{RGB}{111,197,214}
\definecolor{ppu-circle-blue}{RGB}{75,172,198}

\newcommand{\circlearound}[1]{\space\unitlength1ex\begin{picture}(2.5,2.5)%
	\put(0.75,0.75){\circle{2.5}}\put(0.75,0.75){\makebox(0,0){#1}}\end{picture}}

\newcommand*\mandatory{\tikz[baseline=(char.base)]{
		\node[shape=circle,line width= 0.2mm, fill=vat-common-green, draw=vat-common-green, text=white, inner sep=1.1pt] (char) {!};}}

\newcommand*\optional{\tikz[baseline=(char.base)]{
		\node[shape=circle,line width= 0.2mm, fill=vat-optional-blue, draw=vat-optional-blue, text=white, inner sep=1.1pt] (char) {?};}}

\newcommand*\alternative{\tikz[baseline=(char.base)]{
		\draw[-{Triangle[width=8pt,length=4pt, color=vat-alternative-yellow]}, line width=2pt, color=vat-alternative-yellow](0.2,-0.05) -- (0.6,-0.05);
		\draw[-{Triangle[width=8pt,length=4pt, color=vat-alternative-blue]}, line width=2pt, color= vat-alternative-blue](0.4,0.05) -- (0.0, 0.05);}}

\modulolinenumbers[5]
\definecolor{keyword_color}{rgb}{0,0,205}
\lstdefinelanguage{PLCOpen}{
	keywords={IF, THEN, END\_IF, PROGRAM, VAR, END\_VAR,BOOL},
	keywordstyle=\color{keyword_color}\ttfamily,
	ndkeywords={FALSE, TRUE},
	ndkeywordstyle=\color{olive}\ttfamily,
	identifierstyle=\color{black},
	sensitive=false,
	comment=[l]{//},
	morecomment=[s]{/*}{*/},
	commentstyle=\color{purple}\ttfamily,
	stringstyle=\color{red}\ttfamily,
	morestring=[b]',
	morestring=[b]"
}

\lstdefinelanguage{ST}{
	keywords={LD, AND, OR, ST, IF, THEN, END_IF},
	keywordstyle=\color{blue}\bfseries,
	ndkeywords={class, export, boolean, throw, implements, import, this},
	ndkeywordstyle=\color{darkgray}\bfseries,
	identifierstyle=\color{black},
	sensitive=false,
	comment=[l]{//},
	morecomment=[s]{/*}{*/},
	commentstyle=\color{purple}\ttfamily,
	stringstyle=\color{red}\ttfamily,
	morestring=[b]',
	morestring=[b]"
}

\journal{JSS Special Issue on Software Clones}

\usepackage[nolist]{acronym}

\begin{document}

\begin{frontmatter}

\title{Custom-Tailored Clone Detection for IEC 61131-3 Programming Languages}

\author[tubs-address]{Kamil Rosiak}
\author[tubs-address]{Alexander Schlie}
\author[tubs-address]{Lukas Linsbauer}
\author[tum-address]{Birgit Vogel-Heuser}
\author[tubs-address]{Ina Schaefer}
\address[tubs-address]{Technische Universität Braunschweig}
\address[tum-address]{Technische Universität München}

\tnotetext[t1]{E-Mail addresses: k.rosiak@tu-bs.de (K. Rosiak), a.schlie@tu-bs.de (A. Schlie) , l.linsbauer@tu-bs.de (L. Linsbauer), i.schaefer@tu-bs.de (I. Schaefer), vogel-heuser@tum.de (B. Vogel-Heuser)}

\begin{abstract}
Automated production systems (aPS) are highly customized systems that consist of hardware and software.
Such \acs{aPS} are controlled by a \ac{PLC}, often in accordance with the IEC~61131-3 standard that divides system implementation into so-called \acp{POU} as the smallest software unit and is comprised of multiple textual (\ac{ST}) and graphical (\ac{FBD}, \ac{LD}, and \ac{SFC}) programming languages that can be arbitrarily nested.

A common practice during the development of such systems is reusing implementation artifacts by copying, pasting, and then modifying code. This approach is referred to as code cloning.
It is used on a fine-granular level where a \acp{POU} is cloned within a system variant.
It is also applied on the coarse-granular system level, where the entire system is cloned and adapted to create a system variant, for example for another customer.
This ad hoc practice for the development of variants is commonly referred to as clone-and-own.
It allows the fast development of variants to meet varying customer requirements or altered regulatory guidelines.
However, clone-and-own is a non-sustainable approach and does not scale with an increasing number of variants. 
It has a detrimental effect on the overall quality of a software system, such as the propagation of bugs to other variants, which harms maintenance.

In order to support the effective development and maintenance of such systems, a detailed code clone analysis is required.
On the one hand, an analysis of code clones within a variant (i.e., clone detection in the classical sense) supports experts in refactoring respective code into library components.
On the other hand, an analysis of commonalities and differences between cloned variants (i.e., variability analysis) supports the maintenance and further reuse and facilitates the migration of variants into a \ac{SPL}.

In this paper, we present an approach for the automated detection of code clones within variants (intra variant clone detection) and between variants (inter variant clone detection) of IEC61131-3 control software with arbitrary nesting of both textual and graphical languages.
We provide an implementation of the approach in the \ac{VAT} as a freely available prototype for the analysis of IEC~61131-3 programs.
For the evaluation, we developed a meta-model-based mutation framework to measure our approach's precision and recall.
Besides, we evaluated our approach using the \ac{PPU} and \ac{xPPU} scenarios. 
Results show the usefulness of intra and inter clone detection in the domain of automated production systems.
\end{abstract}

\begin{abstract}
Automated production systems (aPS) are highly customized systems that consist of hardware and software.
Such aPS are controlled by a programmable logic controller (PLC), often in accordance with the IEC~61131-3 standard that divides system implementation into so-called program organization units (POUs) as the smallest software unit and is comprised of multiple textual (Structured Text (ST)) and graphical (Function Block Diagram (FBD), Ladder Diagram (LD), and Sequential Function Chart(SFC)) programming languages that can be arbitrarily nested.

A common practice during the development of such systems is reusing implementation artifacts by copying, pasting, and then modifying code. This approach is referred to as code cloning.
It is used on a fine-granular level where a POU is cloned within a system variant.
It is also applied on the coarse-granular system level, where the entire system is cloned and adapted to create a system variant, for example for another customer.
This ad hoc practice for the development of variants is commonly referred to as clone-and-own.
It allows the fast development of variants to meet varying customer requirements or altered regulatory guidelines.
However, clone-and-own is a non-sustainable approach and does not scale with an increasing number of variants. 
It has a detrimental effect on the overall quality of a software system, such as the propagation of bugs to other variants, which harms maintenance.

In order to support the effective development and maintenance of such systems, a detailed code clone analysis is required.
On the one hand, an analysis of code clones within a variant (i.e., clone detection in the classical sense) supports experts in refactoring respective code into library components.
On the other hand, an analysis of commonalities and differences between cloned variants (i.e., variability analysis) supports the maintenance and further reuse and facilitates the migration of variants into a software productline (SPL).

In this paper, we present an approach for the automated detection of code clones within variants (intra variant clone detection) and between variants (inter variant clone detection) of IEC61131-3 control software with arbitrary nesting of both textual and graphical languages.
We provide an implementation of the approach in the variability analysis toolkit (VAT) as a freely available prototype for the analysis of IEC~61131-3 programs.
For the evaluation, we developed a meta-model-based mutation framework to measure our approach's precision and recall.
Besides, we evaluated our approach using the Pick and Place Unit (PPU) and Extended Pick and Place Unit (xPPU) scenarios. 
Results show the usefulness of intra and inter clone detection in the domain of automated production systems.
\end{abstract}


\end{frontmatter}


\section{Introduction}
\label{sec:introduction}

During the evolution of software systems, code cloning is a common practice~\cite{mondal2020survey} for reusing software artifacts.
To cope with an increasing market for custom-tailored software systems, developers often follow a clone-and-own approach where existing variants are copied and altered to create new variants~\cite{Fischer2014}.
It is an unsustainable approach that reduces the overall software quality due to bug propagation, increases the maintenance effort, and hinders further reuse~\cite{Deissenboeck2010}.
In the field of clone detection, research focuses on high-level programming languages such as Java or C~\cite{mondal2020survey,ain2019systematic,roy2007survey,bellon2007comparison}.
In the domain of \acp{aPS}, code cloning is a common practice due to frequently changing products, customer requirements, and altered regulatory guidelines~\cite{DurdikSustainability,Legat:2013}.

The state of the art programming languages for programming logical controller software is defined in the IEC~61131-3 standard~\cite{IEC61131-3}.
It comprises five programming languages, the two textual languages \acf{ST} and \acf{IL}, and the three graphical languages \acf{SFC}, \acf{LD}, and \acf{FBD}.
The standard allows the nesting of languages languages, such as using \acf{ST} in \acf{FBD} implementations.
The control program developers can select the language that is best suited for a particular task, significantly increasing their productivity.
Programs implemented according to IEC~61131-3 are divided into \acp{POU} as the smallest software unit in a program.
Such systems are often reused by copying the whole system and then modifying it to create new and independent system variants (referred to as clone-and-own).
Furthermore, developers also often reuse single \acp{POU} within a system (referred to as classical code cloning), for example, the POU that controls a sorting conveyor that can occur several times in a production system~\cite{vogel2018maintainability,bougouffa2019visualization}.

To restore the sustainable development of cloned system variants, they need to be re-engineered into a structured reuse approach, such as a \acf{SPL}~\cite{northrop2002software,fischer2018qualitative}.
Therefore, a detailed analysis of system variants concerning code clones within a variant (intra clone detection) and commonalities and differences between cloned variants (inter clone detection) is essential.
It serves as a first step to re-engineer system variants into an \ac{SPL}~\cite{Breivold2008,krueger2001easing} and to refactor code clones into reusable and configurable software artifacts such as library components~\cite{vogel2018key}.

We propose a fully customizable comparison approach for IEC~61131-3 in order to support the detection of clones within a variant (intra variant clone detection) and between variants (inter variant clone detection).
This supports developers in tracing clones within and between variants, which helps them create reusable components within systems and migrating system variants into an \ac{SPL}, respectively.
Specifically, the contributions of this paper are as follows:

\begin{itemize}
	\item A model-based, fine-grained, and fully customizable approach for the detection of code clones within variants (intra clone detection) and analysis of commonalities and differences between cloned variants (inter clone detection) of IEC~61131-3 programs composed of arbitrarily nested sub languages.
	\item Publicly available prototype implementation called \ac{VAT}, evaluation data and results\footnote{\url{https://github.com/TUBS-ISF/IEC_61131\_3\_Clone\_Detection}}.
	\item A mutation framework for the evaluation of clone detection tools for IEC~61131-3 systems.
	\item Detailed evaluation and analysis of the approach by applying it to a large clone data set created using the mutation framework, as well as to the PPU and xPPU case study systems.
\end{itemize}

The remainder of this paper is structured as follows:
\autoref{sec:background} provides relevant background on the IEC~61131-3 standard with the utilized programming languages and describes code clones and variability analysis.
\autoref{sec:concept} presents our approach for detecting clones within and between variants.
In \autoref{sec:implementation_sec}, we explain the implementation of our approach as a tool called \ac{VAT}.
In \autoref{sec:evaluation}, we evaluate our approach by performing qualitative and quantitative analyses.
Finally, we discuss related work in \autoref{sec:related_work} and conclude in \autoref{sec:conclusion}.

\section{Background}
\label{sec:background}
This section provides background on IEC~61131-3 control software, types of code clones, and variability analysis.

\subsection{IEC~61131-3 Control Software} 
Automated production systems are typically controlled by \acp{PLC}, which are typically programmed in accordance with the IEC~61131-3 standard.
A \ac{PLC} executes programs in a cycle that is divided into three phases input scan, program execution, and update of outputs.
\begin{figure}
	\centering
	\includegraphics[width=\linewidth]{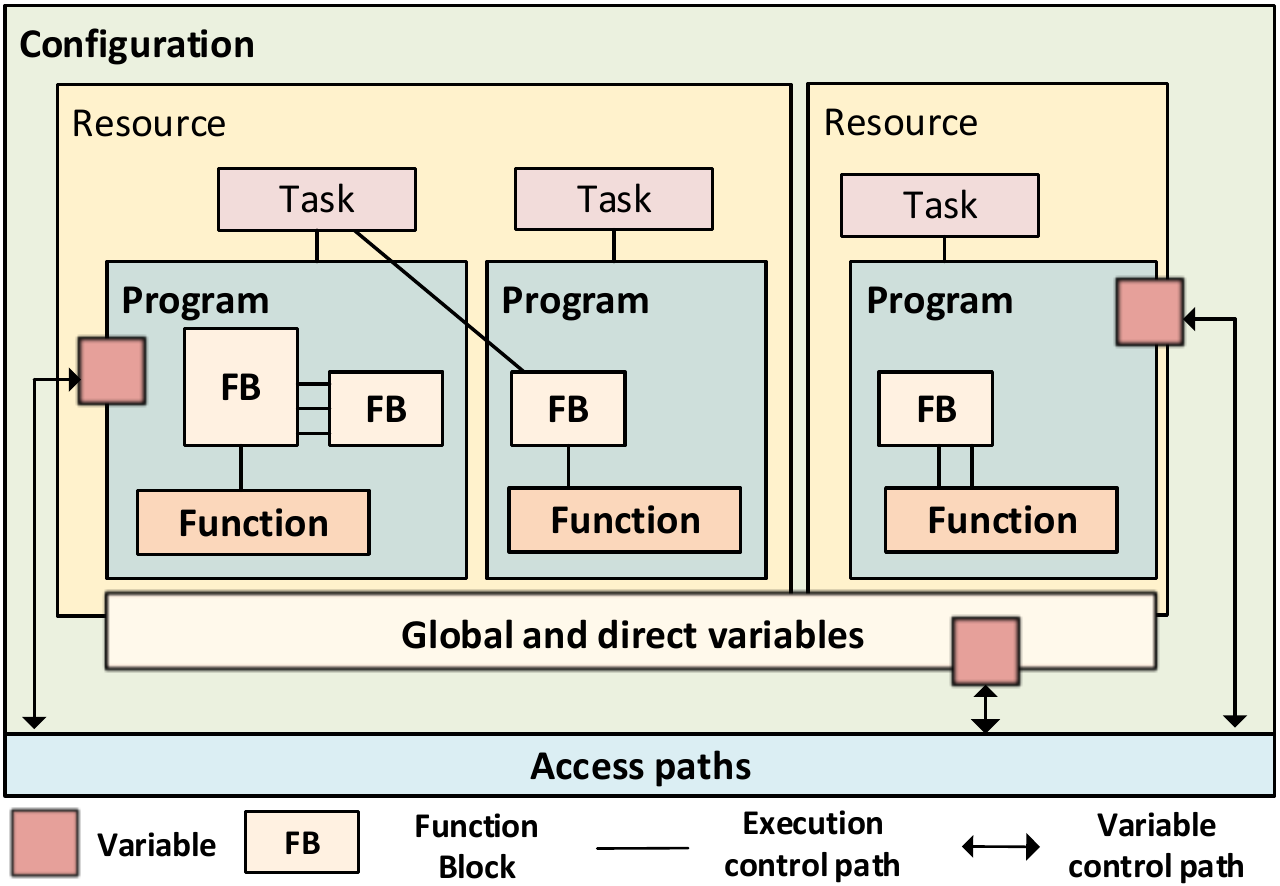}
	\caption{IEC~61131-3 software model~\cite{iec200361131}.}
	\label{fig:IEC61131_software_model}
\end{figure}
Figure~\ref{fig:IEC61131_software_model} illustrates the IEC~61131-3 software model.
The highest level of a \ac{PLC} controlled system is described in a \emph{configuration}, which is assigned to a particular type of control system, including the hardware, i.e., processing resources, memory addresses for all input and output channels, as well as the system capabilities.
Within a \emph{configuration}, there are \emph{resources}, which are a processing facility that can execute IEC programs.
A \emph{resource} can contain one or more \emph{tasks} controlling the execution of \emph{programs} and \emph{function blocks}.
\emph{Programs} are defined as a logical assembly of all programming language elements and constructs necessary to fulfill plant machinery's control task.
\emph{Functions} and \emph{function blocks} are the basic elements and contain specific implementation.
\emph{Functions} do not have a state and always return the identical output given the same input.
In contrast, \emph{function blocks} contain a state and tracking the execution history.
\emph{Programs}, \emph{function blocks} and \emph{functions} are called \acfp{POU}, within the IEC~61131-3.
A \ac{POU} contains a declaration part (\circlearound{1} \autoref{fig:running_example}) where variables and data types are defined and a body part (\circlearound{2} \autoref{fig:running_example}) where algorithms are implemented.
For the implementation of \acp{POU}, the IEC~61131-3 provides multiple languages, which are Structured Text (ST) shown in \autoref{ssec:st}, Function Block Diagram (FDB) reflected in \autoref{ssec:fbd}, Ladder Diagram (LD) illustrated in \autoref{ssec:ld}, Sequential Function Chart (SFC) which is explained in \autoref{ssec:sfc}.
The last implementation language is Instruction List (IL), an assembly-like programming language, which we don't consider due to its deprecation in the previous version of the IEC~61131-3 standard.
The control program developers can select the language that is best suited for a particular task, significantly increasing their productivity. 
Moreover, different languages can be nested so that developers can flexibly switch languages based on the specific tasks, as explained in \autoref{ssec:iec_nesting}.

\subsubsection{Structured Text (ST)}
\label{ssec:st}
\ac{ST} is a high-level textual language that looks syntactically similar to C or Pascal \cite{ramanathan2014iec}.
An \ac{ST} implementation is a composition of single steps called statements (cf.\autoref{st_listing}).
Available statement types are \emph{for}, \emph{while}, \emph{if}, \emph{case}, \emph{assignment}, and \emph{function call}.
\ac{ST} allows implementing complex algorithms, long mathematical functions, array manipulation, and repetitive tasks.

\begin{lstlisting}[language=ST, basicstyle=\scriptsize, frame=single, firstnumber=1, label=st_listing, caption=Example of a Structured Text program., numbers=left]
IF A THEN 
	D := A AND (B OR C);
END_IF
\end{lstlisting}
\autoref{st_listing} shows an example \ac{ST} implementation, which assigns the logical expression $(A \land (B\lor C))$ to the variable D under the condition that A is true.

\subsubsection{Function Block Diagram (FBD)}
\label{ssec:fbd} 
\ac{FBD} comes from the field of signal processing, where integer and/or floating-point values are processed \cite{john2010iec}.
It is a graphical programming language that can describe the function between input and output variables.
A \ac{POU} implemented in \ac{FBD} contains a declaration part and an implementation part. 
The declaration part is used to define variables or constants, either graphical or textual.
The implementation part uses \emph{networks} on the top layer to structure the implementation.
\emph{Networks} represent either a logical or arithmetic expression.
Every \emph{network} has a mark that can be used as a jump target from other \emph{networks}.
For the implementation of \emph{networks}, we can use functions such as logical AND and function blocks such as timer on delay. 
All function blocks have input and output ports, which can be connected with variables or with other blocks. 
Blocks describe a function between input and output.

\begin{figure}[H]
	\centering
	\includegraphics[width=0.6\linewidth]{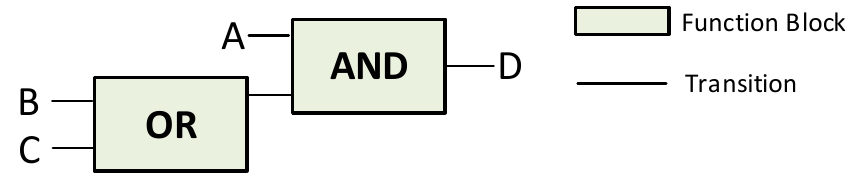}
	\caption{\ac{LD} example}
	\label{fig:fbd_example}
\end{figure}
\autoref{fig:fbd_example} illustrates a \ac{FBD} implementation of a network consisting of a logical AND and a logical OR \emph{Function Block}.
The resulting logical expression of this implementation is $(A \land (B \lor C) = D$.

\subsubsection{Ladder Diagram (LD)}
\label{ssec:ld} 
\begin{figure}[H]
	\centering
	\includegraphics[width=0.5\linewidth]{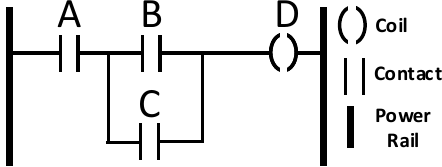}
	\caption{\ac{LD} example}
	\label{fig:ld_example}
\end{figure}

\ac{LD} is a graphical programming language such as FBD and resembles an electric circuit structure.
An LD implementation contains a series of \emph{networks} that are limited on the left and right sides by a current vertical line called the power rail shown in \autoref{fig:ld_example}. 
\emph{Networks} are implemented using \emph{contacts}, \emph{coils}, and connecting lines as in a circuit diagram.
\emph{Contacts} pass the condition true and false from left to the right side.
A \emph{coil} transmits the value of the connections from left to right and stores it in a variable. 
In addition to \emph{contacts} and \emph{coils}, the usage of function blocks and programs is allowed.

\autoref{fig:ld_example} shows an \ac{LD} implementation which uses three contacts (A,B,C) and a coil (D) to express the following expression $(A \land (B \lor C) = D$.

\subsubsection{Sequential Function Chart (SFC)}
\label{ssec:sfc} 
\begin{figure}[H]
	\centering
	\includegraphics[width=0.5\linewidth]{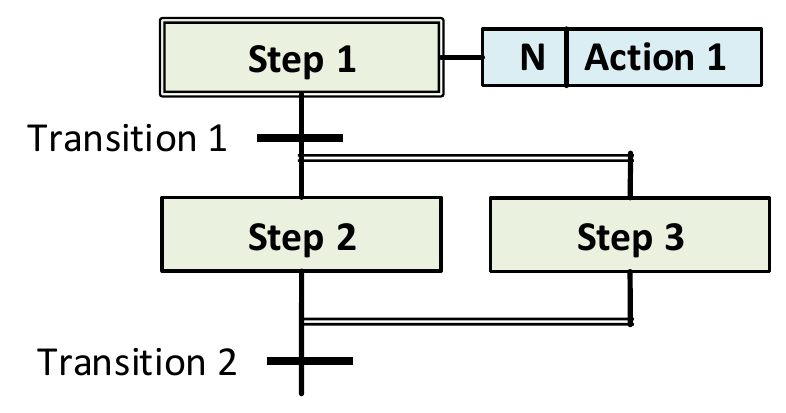}
	\caption{\ac{LD} example}
	\label{fig:sfc_example}
\end{figure}

SFC is a graphical programming language used for PLCs based on binary Petri nets \cite{lewis1998programming}. 
It can be used to program processes divided into single steps. 
The main components of an SFC implementation are Steps with associated Actions and Transitions with assigned conditions, as shown in \autoref{fig:sfc_example}. 
An SFC implementation consists of a series of \emph{steps} connected with directed Transitions. 
\emph{Steps} in an \ac{SFC} implementation can be active or inactive.
When a Step is active, the associated actions are executed. 
A \emph{step} is triggered. 
Either the step is an initial step specified by the developer, or all the steps above are active, and the connecting transition is active. 
Actions can be either entry or exit action.
An entry \emph{action} is executed right after a \emph{step} is activated, and the exit action is executed after the step turns from active to inactive.

\autoref{fig:sfc_example} shows an example \ac{SFC} implementation.
Step1 is the initial step with the assigned Action1.
When the condition of Transition1 evaluates to true, Step2 and Step3 are executed parallel.
Both steps are joined when the Transition2 condition evaluates to true.

\subsection{Nesting of IEC61131-3 Languages}
\label{ssec:iec_nesting}
An additional challenges for the analysis of IEC~61131-3 is the language nesting.
\begin{figure}
	\centering
	\includegraphics[width=\linewidth]{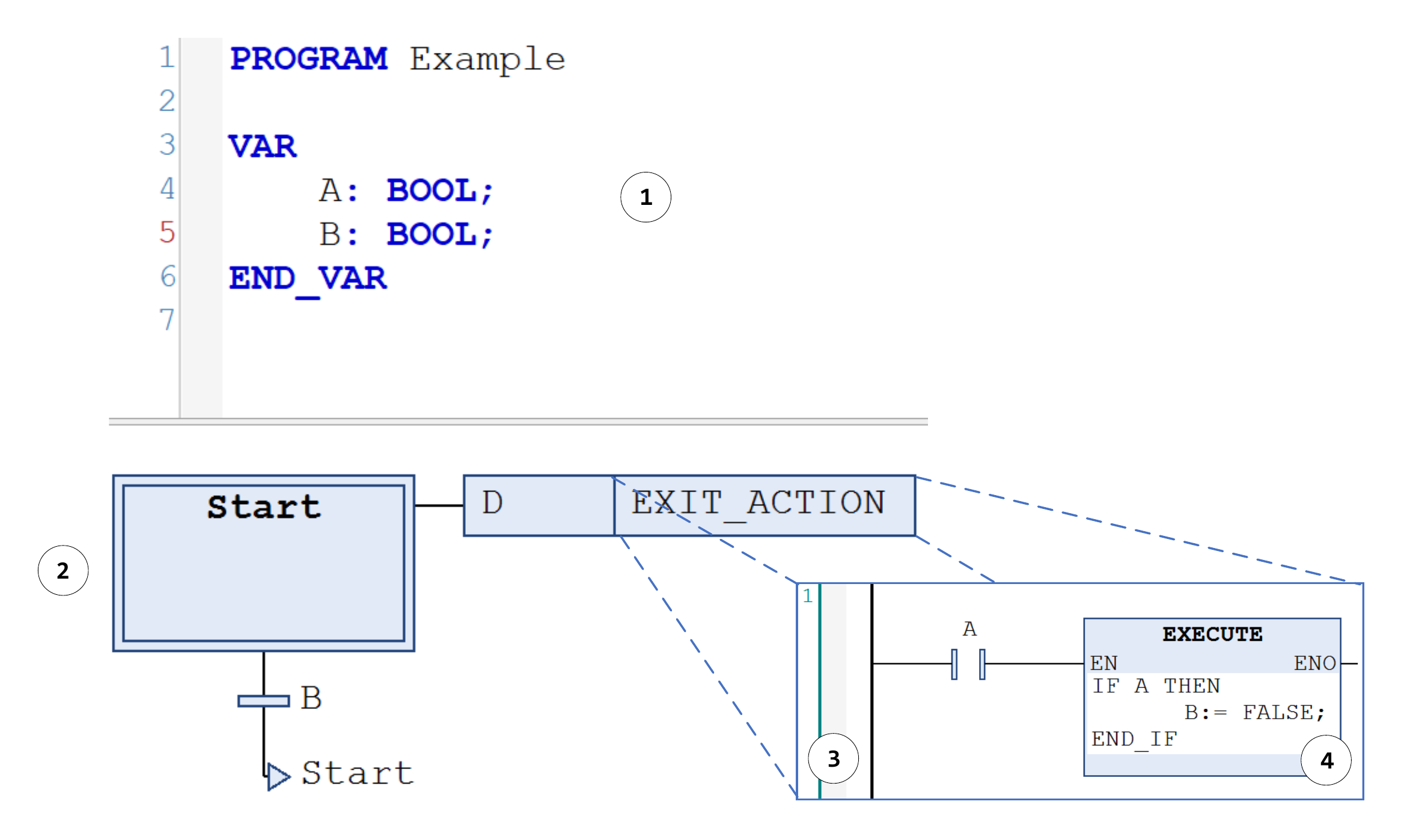}
	\caption{Example IEC61131-3 Program utilizing four programming languages shown in TwinCAT3\cite{TwinCAT}.}
	\label{fig:running_example}
\end{figure}
\autoref{fig:running_example} shows the implementation of a program using four languages for its implementation.
On the top, we can see the global variable declaration with the definition of the Boolean variables A and B \circlearound{1}.
The program is implemented in \ac{SFC} \circlearound{2} and has one step that executes a time-delayed (D) action, which is implemented using \ac{LD}.
In the \ac{LD} implementation \circlearound{3}, we use an execute function block that executes \ac{ST} code \circlearound{4}.

\subsection{Types of Clones}
\label{ssec:code_clone}
Two fragments of code that are similar or even equal are called code clones.
The similarity between code clones can be assessed based either on their textual representation or on their functionality.
Textual clones are often the result of copying and pasting existing code to another location where it can then be adapted if necessary.
Roy et al.~\cite{roy2007survey} classify code clones according to the following four types:
\begin{itemize}
	\item \textit{Type I}: Two code fragments that are similar except for changes in white space or variation in code comments.
	\item \textit{Type II}: Code fragments that are syntactically equal but can show renaming of literals or identifiers as well as the changes of \textit{Type I} clones.
	\item \textit{Type III}: In addition to the properties of Type 2 clones, the fragments can show further modifications such as additional statements, added, or deleted lines.
	\item \textit{Type IV}: Two code fragments that offer the same functionality but are implemented differently. Type IV clones are also called semantic clones.
\end{itemize}

\begin{figure}[h]
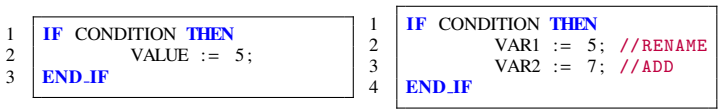

\begin{minipage}{.45\linewidth}
	\begin{lstlisting}[language=ST, basicstyle=\scriptsize, frame=single, firstnumber=1, label=st_listing_1, numbers=left]
IF CONDITION THEN
	VALUE := 5;
END_IF
	\end{lstlisting}
\end{minipage}\hfill
\begin{minipage}{.45\linewidth}
	\begin{lstlisting}[language=ST, basicstyle=\scriptsize, frame=single, firstnumber=1, label=st_listing_2, numbers=left]
IF CONDITION THEN
	VAR1 := 5; //RENAME
	VAR2 := 7; //ADD
END_IF
	\end{lstlisting}
\end{minipage}
\caption{Example Structured Text Clone Pair}
\label{fig:example_st_clone_pair}
\end{figure}

In \autoref{fig:example_st_clone_pair} we show a cloned pair of \ac{ST} implementations.
On the left, if the variable CONDITION evaluates to true (cf. line 1), we assign the value 5 to the variable VALUE (cf. line 2).
On the right, the variable VALUE is renamed to VAR1 (cf. line 2), and the if condition block is extended with an additional assignment (cf. line 3).
So these code fragments are type III clones of each other.
We can also apply the commonly available definition of clones to IEC~61131-3 languages.
The only difference is the name of the artifacts. In the context of IEC systems, we speak of configurations or POU, whereas in object-oriented languages, for example, classes and methods are common artifacts.

\subsection{Variability Analysis} 
\label{sec:variability}
In contrast to clone detection, variability analysis describes the identification of similar and variable parts between a family of software variants.
A family of software is a set of programs that has common properties~\cite{parnas1976design}.
One member of this software family represents a valid realization of one product known as a software variant.
Depending on how such a software family has been created, the effort of creating new variants and maintaining existing products increases rapidly~\cite{northrop2002software}.
A variability analysis is recommended to re-instantiate software variants' sustainable development in a system created with clone-and-own.
Such an analysis can support developers in maintaining software families, e.g., tracing bugs between variants, or helping them by migrating the whole software family into an \ac{SPL}~\cite{northrop2002software,parmeza2015cost,alam2016comprehensive}.

\section{Clone Detection Approach}    	 
\label{sec:concept}
This section presents our approach for the detection of code clones in IEC~61131-3 control software.
We first explain the general comparison approach and then each step in more detail in the following sections.
\begin{figure*}[t]
	\centering
	\includegraphics[width=\linewidth]{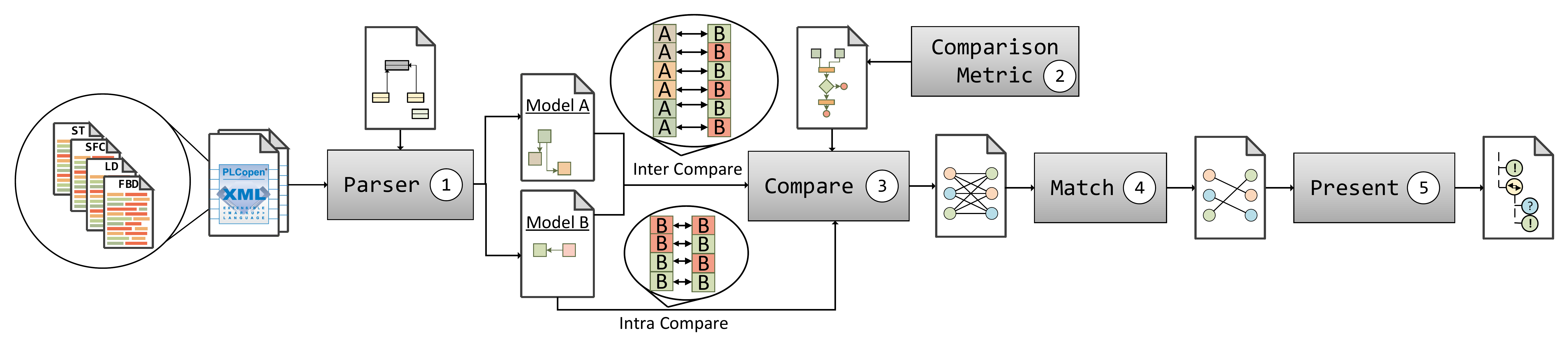}
	\caption{Process for the detection of code clones.}
	\label{fig:process}
\end{figure*}
\autoref{fig:process} illustrates the process for the detection of code clones.

In the first step, the control software is \emph{parsed}~\circlearound{1}.
The parsing process transforms a PLCOpenXML file into a model based on a set of meta-models.
We created these meta-models as an abstraction of the IEC~61131-3 standard to reduce the complexity.
A PLCOpenXML file contains the implementation of a system as a set of \acp{POU}.
Every \ac{POU} may be implemented in one of five languages proposed by the IEC61131-3 and can contain nested implementations in different languages.

A \emph{configurable comparison metric}~\circlearound{2} can customize the comparison process.
This metric is a composition of options and weighted attributes that the user can customize.
The input model(s) are decomposed into smaller elements based on selected options. These elements are then compared with each other, and a similarity value is computed based on the weighted attributes.

The \emph{comparison step}~\circlearound{3} receives as input either one instance of a model representing a system (in case of intra clone detection) or two instances, each representing a system variant (in case of inter clone detection).
The \emph{comparison process} decomposes the input models into smaller elements and then compares them to compute pairwise similarities based on the comparison metric.
The difference between intra and inter clone detection is only in the pairing of elements for comparison.
In the case of intra clone detection (i.e., the detection of clones within a system), all pairwise combinations of all \acp{POU} in one model are created and compared.
In contrast, the inter clone detection (i.e., the detection of clones between variants of a system) receives two models (each representing a variant) as input, and all \acp{POU} within the first model are compared against all \acp{POU} of the second model.
The result of the comparison process is a similarity tree in which each node is a description of a weighted edge in a completed, weighted, bipartite graph.

The \emph{matching step}~\circlearound{4} calculates an independent edge set on this graph to filter it and obtain the most similar elements.
The independent edge set contains information about the relation of elements that are finally \emph{presented}~\circlearound{5} to the user in the form of a family model for more straightforward interpretation than just a list of matched pairs.

\subsection{Parsing Nested IEC~61131-3 Code}
The parsing process transforms a PLCOpenXML file into an instance of our meta-models.
We decided on this file format because most of the tools that deal with IEC~61131-3 code support the export of projects into this format.
Besides, this format reflects the project structure and delivers meta information of the project.
To process the project structure, we created a set of meta-models that capture the extent of PLCOpen projects.

\begin{figure}[h]
	\centering
	\includegraphics[width=0.9\linewidth]{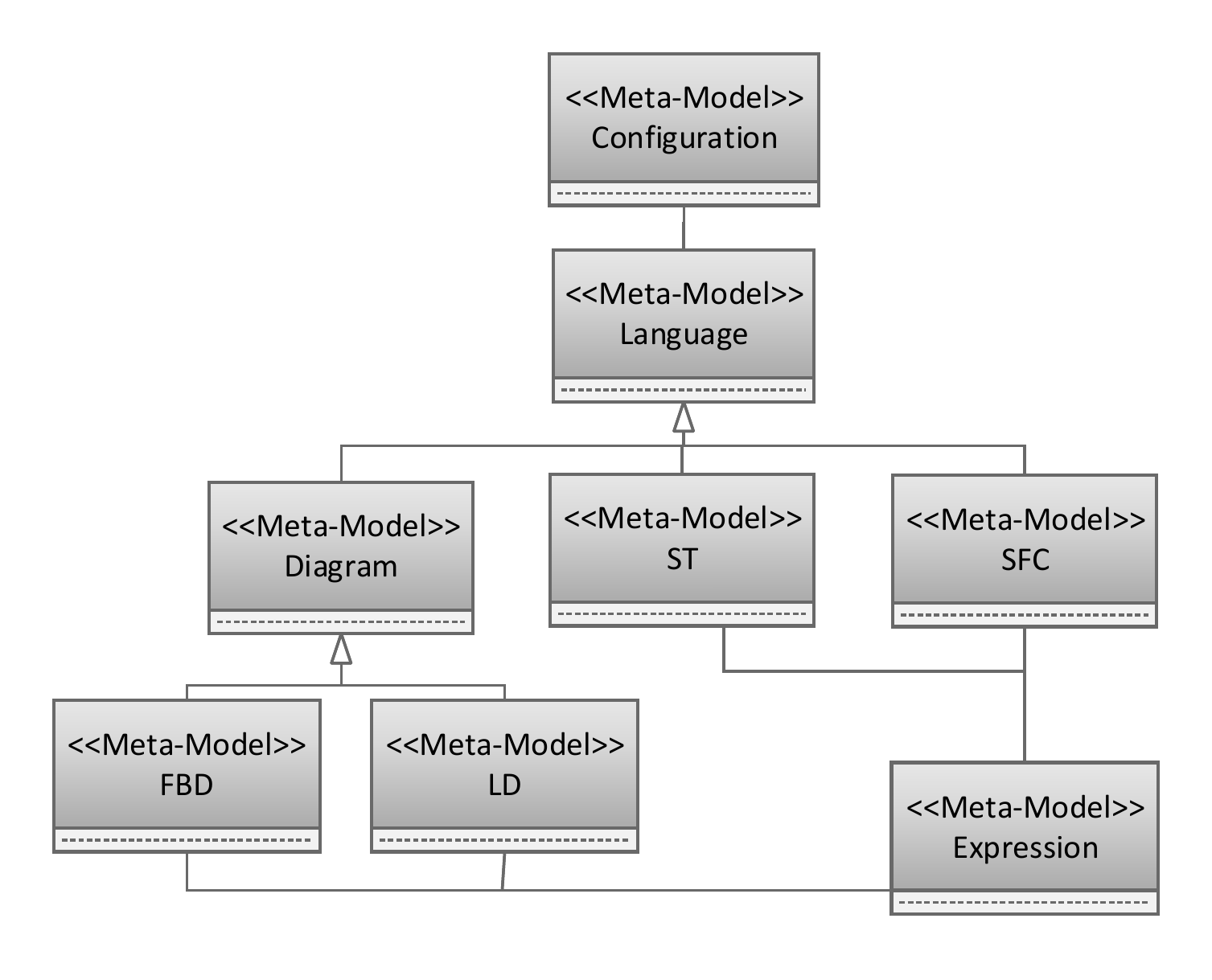}
	\caption{Meta-models and their dependencies as \ac{UML} class diagram.}
	\label{fig:meta_model_dependencies}
\end{figure}
\autoref{fig:meta_model_dependencies} illustrates a schematic overview of our meta-model architecture.
On the top level, the \emph{configuration} meta-model models the project structure and contains elements such as the \acp{POU} and their implementations.
As an abstraction layer between the \emph{configuration} and the implementation of each \ac{POU}, we introduced the \emph{Language} meta-model.
Additionally, we created a meta-model for each of the IEC~61131-3 programming languages that inherit from the Language meta-model, except for \ac{IL}, which has been deprecated in the last version of the standard.
As another abstraction level, the \emph{diagram} meta-model contains the common elements of the two graphical languages \ac{FBD} and \ac{LD} such as networks, ports, and jumps.
Moreover, we created a meta-model for expressions used in all programming languages, such as the condition of contacts in an \ac{LD} implementation or a Boolean expression in an \ac{ST} statement.
We created eight meta-models that contain a total of 51 classes and 18 enumerations.
Generally, our meta-models and their classes allow capturing detailed information about systems and artifacts, which is essential for a detailed analysis of such systems.
Furthermore, our meta-models are an abstraction of the IEC 61131-3 standard, which reduces the complexity of systems implemented in accordance with it.
As input for the parsing process, we utilize a PLCOpenXML file and get a meta-model instance as output, which is the input for the comparison process.

\lstset{emptylines=2}
\begin{figure}[h]
	\begin{minipage}{.45\linewidth}
		\begin{lstlisting}[label=example_pou_declaration,caption=POU declaration,language=PLCOpen]{Name}
PROGRAM EXAMPLE
VAR
	A: BOOL;
	B: BOOL;
END_VAR
		\end{lstlisting}
	\end{minipage}\hfill
	\begin{minipage}{.45\linewidth}
		\begin{lstlisting}[label=example_pou_implementation,caption=POU implementation,language=PLCOpen]{Name}
IF A THEN 
	B:= FALSE;
END_IF
		\end{lstlisting}	
	\end{minipage}
	\includegraphics[width=\linewidth]{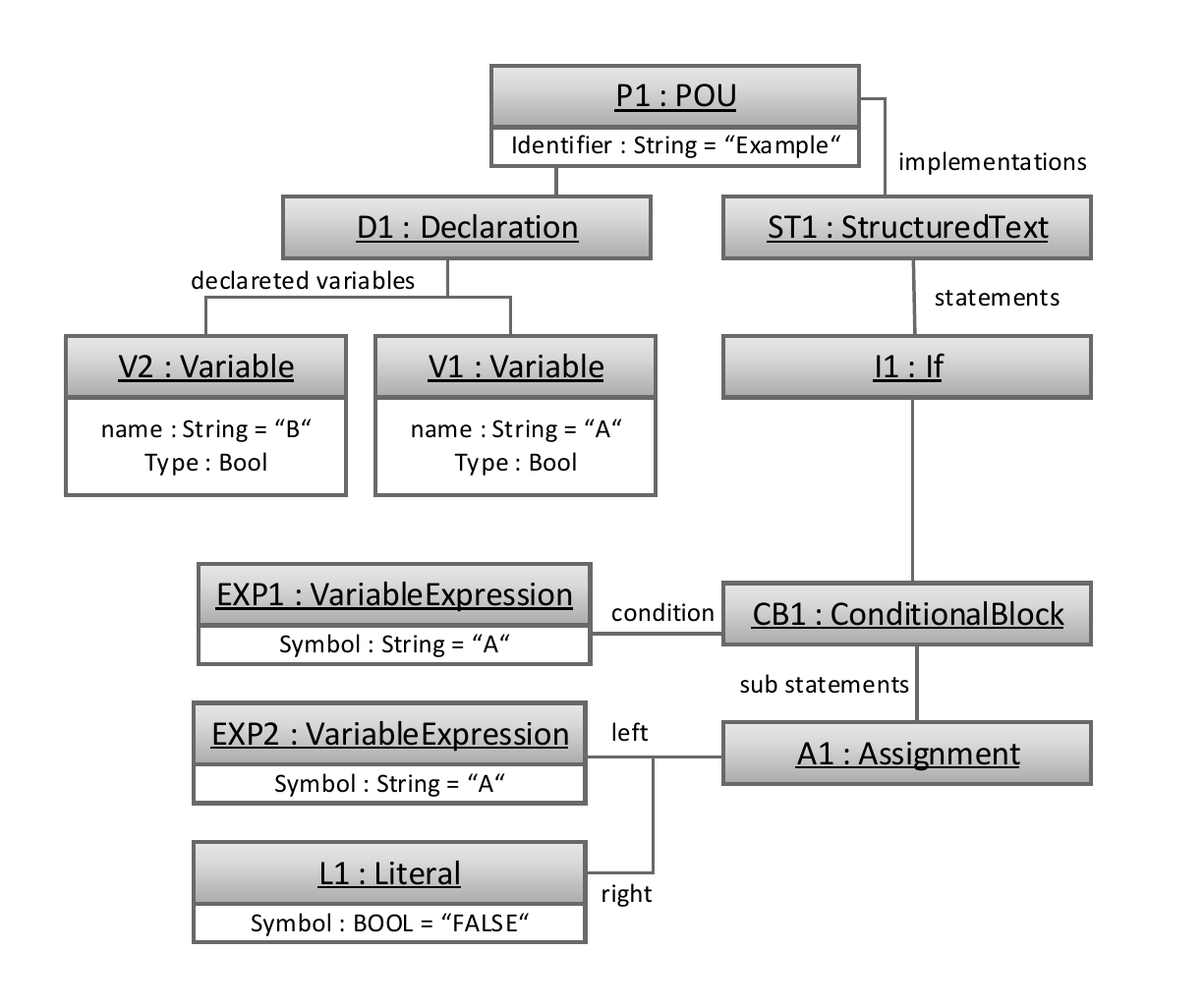}
	\caption{Instance of the meta-model, as \ac{UML} object diagram, representing the IEC~61131-3 code shown in \autoref{example_pou_declaration} and \autoref{example_pou_implementation}.}
	\label{fig:pou_impl}
\end{figure}
In \autoref{example_pou_declaration} and \autoref{example_pou_implementation} we illustrate an example \ac{POU} declaration and implementation, respectively.
\autoref{fig:pou_impl} shows the respective model representation as \ac{UML} object diagram.
The root is the \ac{POU}, which has the \emph{declaration} and the \emph{StructuredText} objects as child elements.
The \emph{declaration} object contains the defined variables A and B and the \emph{implementation} object the statements of our example \ac{ST} implementation.

\subsection{Comparison Metric Definition}
A fully customizable, fine-grained comparison metric drives the comparison of IEC~61131-3 artifacts.
It allows domain experts to customize the comparison process.
A metric is a hierarchical composition of \emph{options} and \emph{attributes}.
\emph{Attributes} are atomic comparison operations that compare two elements with each other, such as comparing two statement types.
Each attribute's result is a float value between 0 and 1, which indicates how similar are the compared elements in percent.
\emph{Options} can activate and deactivate parts of the comparison process, such as the comparison of global variables.
Moreover, it is possible to adjust every attribute and option with a weight, allowing the customization of the metric and prioritization of parts of the comparison.
In general, options define which elements to compare, and attributes define how to compare these elements.

\begin{figure}[h]
	\centering
	\includegraphics[width=0.9\linewidth]{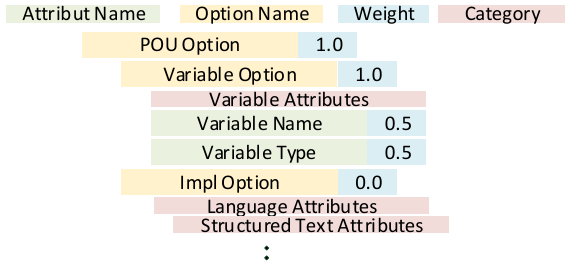}
	\caption{An illustration of a part of our comparison metric.}
	\label{fig:metric_mockup}
\end{figure}
\autoref{fig:metric_mockup} shows a part of a metric for the comparison of \acp{POU}.
The composition of options and attributes has the same structure as the model (cf. \autoref{fig:pou_impl}).
In this graph, options can contain options and attributes, and attributes are leaves.
In the example metric in \autoref{fig:metric_mockup} we compare \acp{POU} only based on their variables (indicated by the X on the left of the respective option) and ignore their implementation.
Variables are compared with two attributes, which compare the type and the name of variables.
Selected attributes are adjusted with a weight that is shown on their right.
In total, our clone detection technique for IEC~61131-3 supports 17 options and 65 attributes for the definition of a custom comparison metric.
A detailed list of all attributes and options is available in our online material \footnote{\url{github.com/TUBS-ISF/IEC\_61131\_3\_Clone\_Detection}}.


More formally, a \emph{comparison metric} $M$ is represented by a root option.
An \emph{option} $o$ is a triple $(O',A,t,w)$ where $O'$ is a set of sub-options, $A$ is a set of attributes, $t$ is the type of artifact the option applies to, and $w \in [0,1]$ is a weight value.
An \emph{Attribute} $a \in o.A$ is a pair $(f,w)$, where $f$ is a function $f(x,y) = s$ with $x$ and $y$ being implementation artifacts (i.e., elements of our meta-model) and $s \in [0,1]$ the computed similarity value, and $w \in [0,1]$ is a weight value.




\subsection{Comparison Process for IEC~61131-3 Models}
In this section, we present the general comparison approach.
In \autoref{ssec:general_comparison} we describe the comparison of model elements and the resulting similarity tree.
Each node of the tree describes a relation between an element of the first and the second input model.
Finally, we describe how we cope with nested implementation artifacts shown in \autoref{ssec:nested_impl}.

\subsubsection{Comparison Approach and resulting Similarity Tree}
\label{ssec:general_comparison}

\begin{algorithm*}
	\begin{algorithmic}[1]
		\Function{compare}{Artifact x, Artifact y, Option o} \Comment{Compare the current pair of artifacts x and y using option o}
		
			\State ArtifactPairSimilarityNode n $\gets (x,y,0.0)$ \Comment{Create Artifact Pair Node $(\text{artifact}_1, \text{artifact}_2, \text{similarity})$}
	
			\ForAll{Option $o' \in o.O'$} \Comment{For all options}
				\State OptionSimilarityNode on $\gets (o', 0.0)$ \Comment{Create Option Node $(\text{option}, \text{similarity})$}
				\ForAll{$(x',y') \in x.\text{children} \times y.\text{children} : x.\text{t} = y.\text{t} = o'.\text{t}$} \Comment{For all pairs of child artifacts that match the option type}
					\State $n'$ $\gets$ \Call{compare}{$x',y',o'$}) \Comment{Recursively compare pairs of child artifacts}
					\State on.children $\gets$ on.children $\cup$ $\{$ $n'$ $\}$ \Comment{Add child artifact pair node as child to current option node}
					\State on.similarity $\gets$ on.similarity + $n'$.similarity \Comment{Add child artifact pair similarity to similarity of current option node}
				\EndFor
				
				\State on.similarity $\gets$ on.similarity $*$ $o'.w$ \Comment{Adjust similarity by option weight}
				
				\State n.children $\gets$ n.children $\cup$ $\{$ on $\}$ \Comment{Add option node as child to current artifact pair node}
				\State n.similarity $\gets$ n.similarity + on.similarity \Comment{Add option node similarity to similarity of current artifact pair node}
			\EndFor
			
			\ForAll{Attribute $a \in o.A$} \Comment{For all attributes}
				\State AttributeSimilarityNode an $\gets (a, a.f(x,y)*a.w)$ \Comment{Create Attribute Node $(\text{attribute}, \text{similarity})$}
				
				\State n.children $\gets$ n.children $\cup$ $\{$ an $\}$ \Comment{Add attribute node as child to current artifact pair node}
				\State n.similarity $\gets$ n.similarity + an.similarity \Comment{Add attribute node similarity to similarity of current artifact pair node}
			\EndFor
	
			\State \Return n

		\EndFunction
	\end{algorithmic}
	\caption{Comparison Algorithm}
	\label{alg:decompose}
\end{algorithm*}

The comparison function shown in Algorithm~\autoref{alg:decompose} receives two artifacts of the same type and the option for comparison.
If both artifacts have child elements and the comparison of these elements is enabled in the metric (i.e., an option with matching type exists), the function is called recursively for each pair of child artifacts.
This way, we systematically perform all pairwise comparisons of elements.
Considering the input as sets of artifacts $X$ and $Y$, we create all pairs $X \times Y$ and compute their similarities based on the comparison metric (specifically, the current option).
The resulting data structure is a tree comprised of three types of similarity nodes: pairs of elements, options, and attributes. 
Pairs of elements can have options and attributes as children, options can only have element pairs as children, and attributes do not have children.

\begin{figure}[h]
	\centering
	\includegraphics[width=\linewidth]{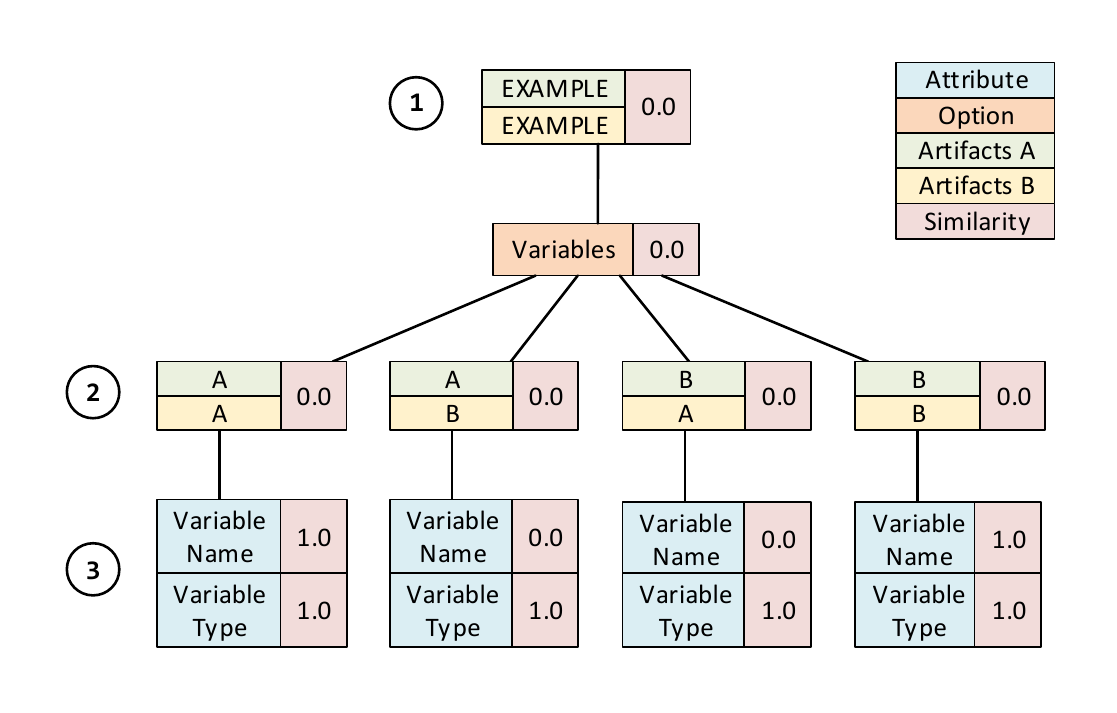}
	\caption{Similarity tree of the comparison of the example model shown in \autoref{fig:pou_impl} with itself using the example metric shown in \autoref{fig:metric_mockup}.}
	\label{fig:solution_structure}
\end{figure}

After the forward path of the recursion in Algorithm~\autoref{alg:decompose}, only the attribute nodes (which are the leaves) of the resulting similarity tree have a similarity value assigned, as shown in \autoref{fig:solution_structure}.
It shows the similarity tree that is created during the comparison of our example model shown in \autoref{fig:pou_impl} driven by the example metric shown in \autoref{fig:metric_mockup}.
Driven by the metric, the process compares both models systematically and creates pairs of elements for the comparison.
The root node of the similarity tree is the pair of \acp{POU} with the name \textit{EXAMPLE} (\circlearound{1} in \autoref{fig:solution_structure}).
Due to the selected \textit{variables} option in the metric (\autoref{fig:metric_mockup}), all pairwise combinations of variables are created and compared.
The pairs are compared using the attributes that are contained in the metric, which are \textit{Variable Name} and \textit{Variable Type} (\circlearound{3} \autoref{fig:solution_structure}).
Only the attributes have a similarity value after the comparison that needs to be propagated to all other nodes.

\begin{figure}[h]
	\centering
	\includegraphics[width=\linewidth]{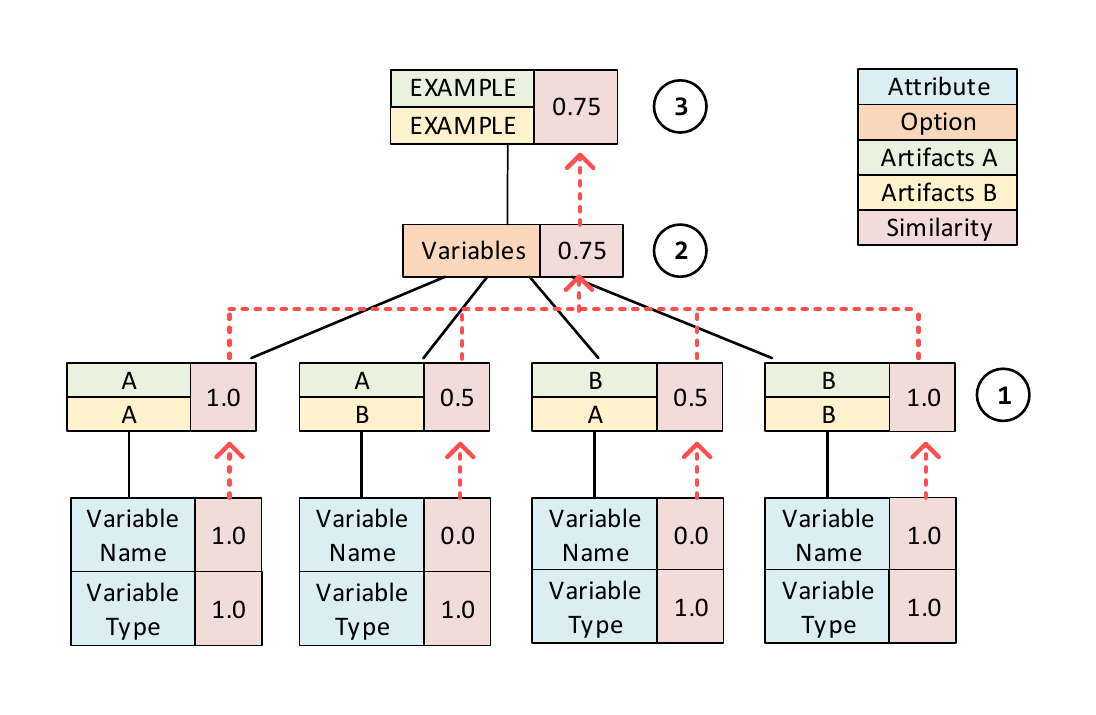}
	\caption{Similarity tree shown in \autoref{fig:solution_structure} after the similarity values have been propagated upward.}
	\label{fig:solution_structure_update}
\end{figure}

The similarity values of the remaining nodes are set on the backward path of the recursion where the weighed similarity values are propagated upward in the similarity tree, as shown in Figure~\ref{fig:solution_structure_update}.
First, the leaf artifact pairs (\circlearound{1} in \autoref{fig:solution_structure_update}) update their similarity by based on the similarities of their attributes.
After that, the \emph{variables} option (\circlearound{2} in \autoref{fig:solution_structure_update}) updates its similarity based on their children's similarity values.
Finally, the similarity value of the root artifact pair (\circlearound{3} in \autoref{fig:solution_structure_update}) is updated.

Each node in this graph describes a weighted edge between an element of $A$ and $B$, which expresses the relation between them (cf. \autoref{fig:solution_structure}.
The edge's weight is a float value between 0 and 1 reflects the similarity between both elements.
All nodes in this graph can be considered as a completed weighted bipartite graph between both sets of elements.

\subsubsection{Detecting Nested Implementations}
\label{ssec:nested_impl}
Our meta-models are designed to store different implementation languages in one model.
This allows us to compare different combinations of nestings in the implementation, such as an action in an \ac{SFC} implementation that is implemented in another language.
To compare different language nestings, we extended the metric with pointers at any place where IEC implementations can be nested. This allows to jump to the corresponding language options and attributes whenever an artifact with a type is detected that corresponds to another language.
In \autoref{tab:base_metric} and \autoref{tab:ppu_metric}, this is indicated by black arrows on the left side.
The whole process is called recursively, which allows comparing nestings of any level. 
We extended the similarity tree with an implementation option for artifacts with a nested implementation.
To the best of our knowledge, this is the only work that provides this functionality.
This comparison is possible based on the abstraction level in our meta-model structure, which allows us to model abstract languages on the one hand. 
On the other hand, the recursively defined comparison approach compares models based on a fully customizable metric.

\subsection{Matching Process using the Similarity Tree}
\begin{figure}[h]
	\centering
	\includegraphics[width=0.8\linewidth]{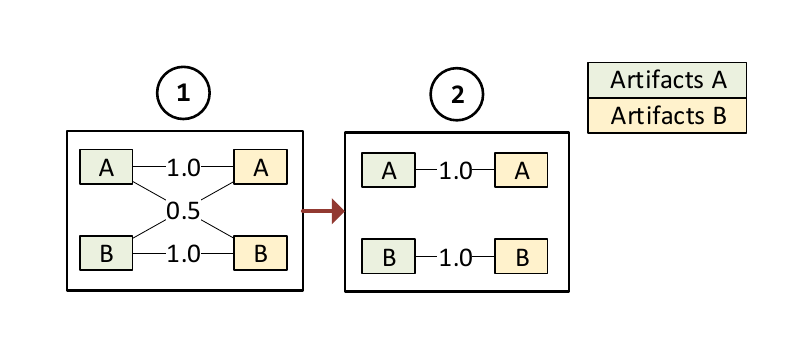}
	\caption{Example completed weighted bipartite graph \circlearound{1} and a calculated independent edge set of this graph \circlearound{2}}
	\label{fig:matching}
\end{figure}

The resulting similarity tree is processed to match each element of the one input model to at most one element of the other input model.
This matching is important, as the relations identified during the comparison are usually ambiguous as elements can be contained in multiple pairs.
For instance, the children of the \emph{variables} option (\circlearound{2} in \autoref{fig:solution_structure_update}) are all pairwise combinations of the input models' variables.
The result contains the triples $(A.A,B.A,1.0)$, $(A.A,B.B,0.5)$, $(A.B,B.A,0.5)$, $(A.B,B.B,1.0)$, which are visualized as a completed, weighted, bipartite graph (\circlearound{1} in \autoref{fig:matching}).
Every node on the left is connected with all nodes on the right and vise versa.
During the matching phase, the graph is filtered and afterward contains only element-pairs with the highest similarity, also known as an independent edge set.
Therefore, we use an approximation algorithm that sorts the edge set by similarity and picks the elements with the highest similarity.
After that, selected elements are marked to prevent them from being selected again.
\circlearound{2} in \autoref{fig:matching} shows the calculated matching of our example.
Finally, we can update the solution and get a similarity of 100\% between both models as expected because we compared a model with itself.
\subsection{Presentation of the Result}
The last step in our process is the presentation.
In contrast to other clone detection approaches that present results as and edit script, we visualize the results as a family model~\cite{beuche2004variants}.
Family models represent the variable architecture of product lines independently of the programming or modeling languages. 
It is a comprehensible representation of commonalities and differences between artifacts.
A further advantage of a family model is that it is possible to derive a domain-level feature model from it.
A family model contains three different element types: mandatory, optional, and alternative are assigned using an adjustable threshold of $\lambda$.
Mandatory elements are marked with an exclamation mark \mandatory{} and have a similarity value larger than $\lambda$, which means that they are common in both models or code clones.
Optional elements are marked with a question mark \optional{} and have a similarity of zero, which means that they are only contained in either of the two variants or code fragments.
The last element type is alternative, which is marked with a left and right arrow \alternative{} and assigned when $0 < similarity < \lambda$. Alternative elements are similar to each other but not equal, as is the case after a smaller code modification.
\begin{figure}[h]
	\centering
	\includegraphics[width=0.85\linewidth]{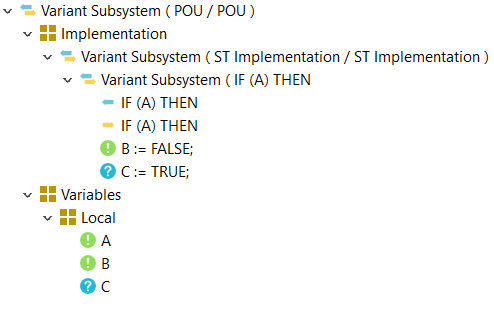}
	\caption{Resulting family model of the comparison between the example in \autoref{fig:pou_impl} and a variant of it.}
	\label{fig:family_model}
\end{figure}
In \autoref{fig:family_model}, we show the resulting family model of the comparison between our example program (\autoref{fig:pou_impl}) and a modification of it.
We extended the implementation with an assignment and added the variable C which are marked with the blue question mark.

\subsection{Summary}
In this section, we showed our approach for the detection of fine-grained changes between implementation artifacts.
This approach can detect code clones in variants and detect variability between variants.
Based on the parsing process, Type I clones are detectable due to the code normalization utilized by the lexer.
Type II clones can be detected based on utilized attributes that compare different properties of the respective implementation artifacts, such as the name of the type or more specific attributes.
The detection of Type III clones in fine-grained elements such as a statement with a changed expression can be detected with attributes.
Cloned code fragments are visualized in the family model with the \alternative{}.
These elements result from the similarities of their nested artifacts and can show changes on cloned code and differences between two variants.
This clone detection can be used to detect variation points between two variants.

\section{Implementation} 			
\label{sec:implementation_sec} 		
In order to evaluate our approach, we implemented it in a publicly available tool we call the \acf{VAT}\footnote{\url{https://github.com/TUBS-ISF/IEC\_61131\_3\_Clone\_Detection}}.

\subsection{Implementation Techniques}
Meta models that describe the IEC~61131-3 project structure are created with the \ac{EMF}, which utilizes Ecore.
Ecore is a meta-model that represents an implementation of the \ac{EMOF}, which is a subset of the \ac{MOF}\footnote{www.omg.org/mof}, a modeling standard defined by the \ac{OMG}\footnote{www.omg.org}.
To transform PLCOpenXML into a model representation, we created a parser using \ac{ANTLR}\footnote{www.antlr.org}, which generates a lexer and a parser based on a grammar file.

The data structures, as well as the compare engine, are implemented using Java 8.
The prototype is developed as \ac{RCP} client based on the Eclipse 4.0\footnote{www.eclipse.org} framework and runs on Windows\footnote{www.microsoft.com} with a \ac{JVM}.
All parts of the \ac{VAT} are plug-ins that are reusable in other applications.

All meta-models, grammar files, and plug-ins with their source code are freely available in our online materials\footnote{\url{https://github.com/TUBS-ISF/IEC\_61131\_3\_Clone\_Detection}}.

\subsection{The Variability Analysis Toolkit}
\label{sec:implementation}
The \ac{VAT} is a tool that supports domain experts with the analysis of IEC~61131-3 programs.
In order to improve the usability of our approach, we implemented a graphical user interface that supports domain experts during every step of our approach.

\subsubsection{Metric Definition}
\begin{figure}[h]
	\centering
	\includegraphics[width=\linewidth]{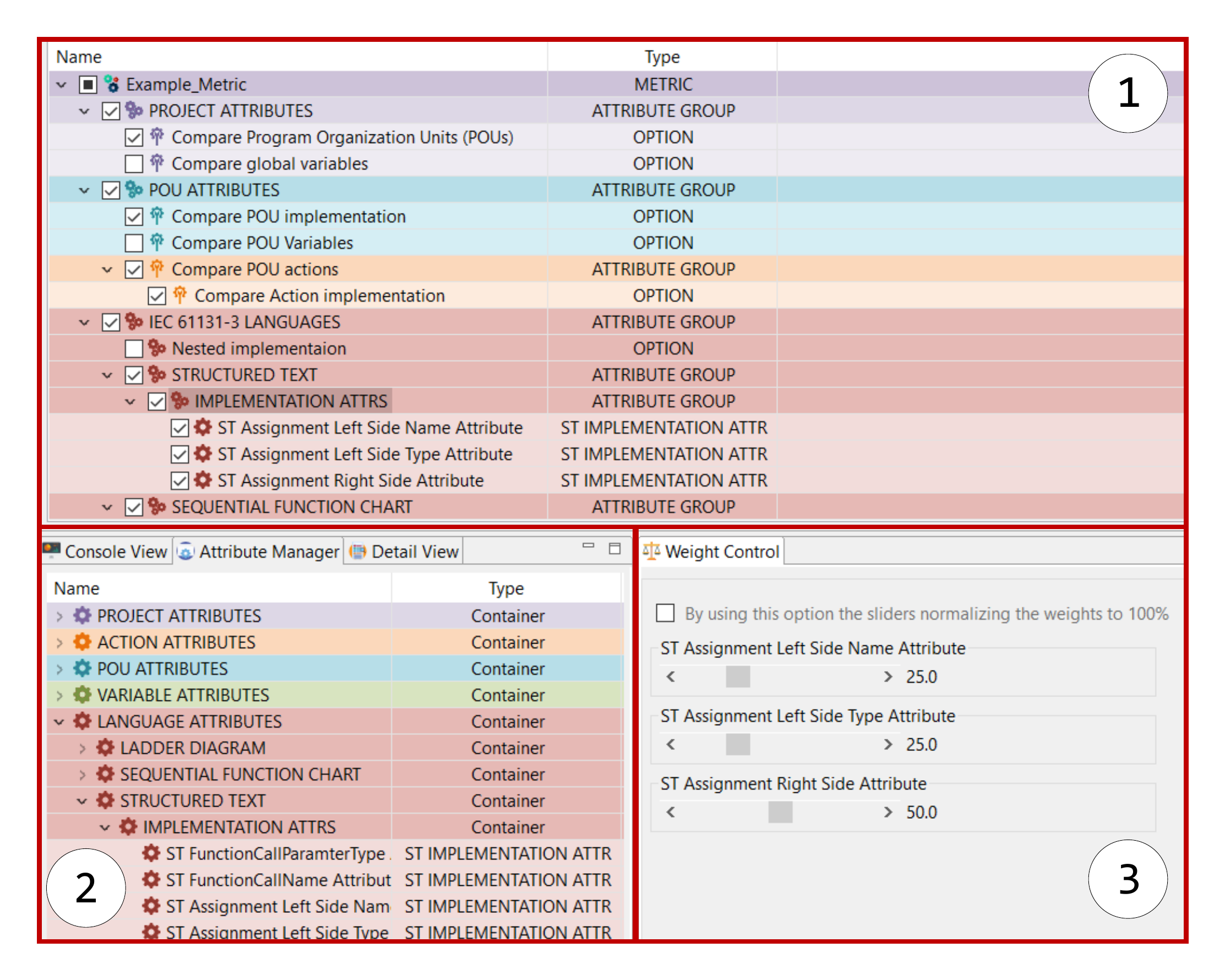}
	\caption{User interface for the metric definition process.}
	\label{fig:vat_gui_metric}
\end{figure}

A screenshot of the interface for the definition of the comparison metric is shown in \autoref{fig:vat_gui_metric}.
The metric definition is supported by the Metric Manager \circlearound{1}.
In the Metric Manager, we can see the base structure of a metric, which contains options that allow us to enable and disable parts of the comparison approach.
Attributes can be selected in the Attribute Manager \circlearound{2}, which contains 66 predefined attributes that are sorted by artifact categories.
For the comparison of fine-grained artifacts we implemented 29 attributes in total, that can compare fine-grained implementation artifacts such as single statements or function blocks.
We implemented 11 attributes for the comparison of \ac{ST}, 6 attributes for \ac{SFC}, 5 attributes for \ac{LD}, and 7 attributes for \ac{FBD}.
To adjust the weights of options and attributes, which express their impact on the comparison process, we can use the Weight Controller \circlearound{3}.
Metrics can be stored and exchanged, which allows domain experts to discuss utilized metrics.

\subsubsection{Comparison and Presentation}
\begin{figure}[ht!]
	\centering
	\includegraphics[width=\linewidth]{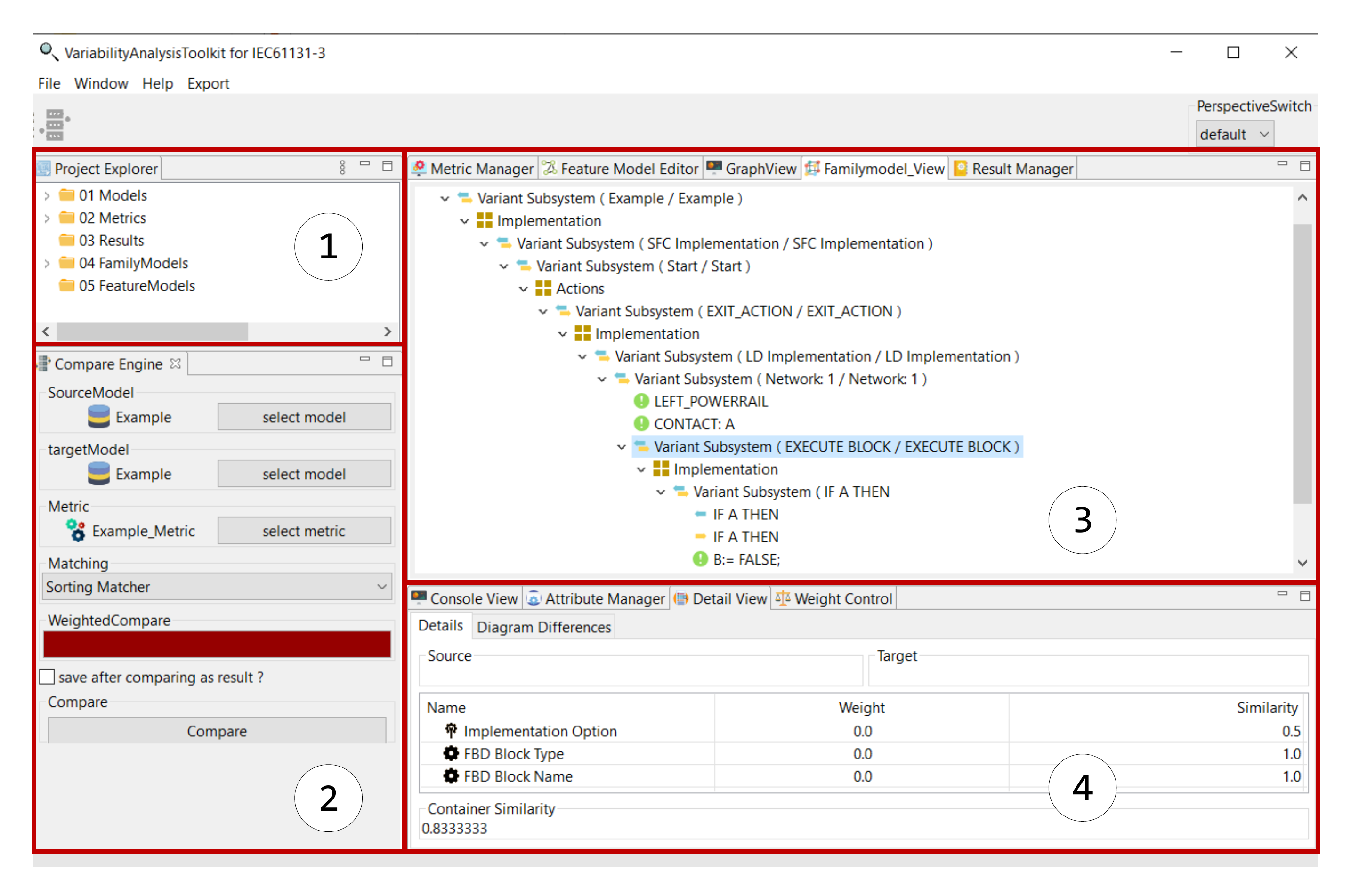}
	\caption{User interface for the comparison and presentation process.}
	\label{fig:vat_gui_compare}
\end{figure}

A screenshot of the comparison process and the presentation of the results is sown in \autoref{fig:vat_gui_compare}.
To start the comparison approach, at least one model from the project explorer \circlearound{1} must be selected.
In a context menu, the user can then select if an intra or inter variant clone detection shall be performed, which changes the selected models' decomposition process.
After selecting the comparison mode, the compare engine \circlearound{2} shows, and the user can select a metric and decide if it wants to compare weighted or not.

After the comparison, the results are presented in family model \circlearound{3}.
The comparison details, such as element similarities per attribute, can be inspected using the Detail View  \circlearound{4}.
This view shows utilized attributes, options, and their weights as well as the resulting similarity value and supports experts in the comprehension of the created results.

\section{Evaluation} 	 			
\label{sec:evaluation}				
We evaluated different aspects of our clone detection approach.
The correctness, measured in precision and recall, of results are crucial for detecting code clones within software variants and analyzing commonalities and differences between software variants.
Otherwise, incorrectly matched elements inevitably compromise subsequent steps such as refactoring code clones into library components or consolidating a set of variants into an \ac{SPL}. 
Thus, analyzing the results concerning their correctness is an essential goal of our evaluation.
Another key aspect for the productive use of the toolkit is its scalability in terms of run time and memory consumption.

\subsection{Research Questions}
\label{ssec:rq} 
Based on the requirements we identified to be relevant, we defined the following research questions (RQs) that we address in this evaluation:
\\
\noindent
\textit{RQ1 Correctness:} With what precision and recall can we detect code clones in arbitrarily nested IEC~61131-3 programming languages?
\begin{itemize}
	\item \textit{RQ 1.1:} What is the impact of different comparison metrics on precision and recall?
\end{itemize}
\textit{RQ2 Scalability:} Does the clone detection approach scale?
\begin{itemize}
	\item \textit{RQ 2.1:} Is the run time of the comparison approach within reasonable bounds? 
	\item \textit{RQ 2.2:} Is the memory consumption of the comparison approach within reasonable bounds? 
\end{itemize}
\textit{RQ3 Usefulness:} How useful is the approach when applied to realistic industrial subject systems?
\begin{itemize}
	\item \textit{RQ 3.1:} How similar are the different \ac{PPU} and \ac{xPPU} scenarios to each other based on a fine-grained and coarse-grained metric?
	\item \textit{RQ 3.2:} How many clones per type can be identified during the evolution of the PPU and xPPU scenarios?
\end{itemize}

\subsection{Setup} 
We evaluated our approach on an Intel Core i7-3770k (3,5 GHz) with 16 GB of RAM, running Windows 10 64bit.
We utilize two different comparison metrics.
On the one hand, we use a coarse-grained metric that compares the implementation using attributes that count the number of specific artifacts, such as how many steps are in an \ac{SFC} implementation. 
On the other hand, we use a fine-grained metric to detect fine-grained changes between single statements, such as an extended condition.
This lets us draw conclusions on the impact of a fine-grained comparison metric in comparison to a coarse-grained comparison metric.
All weights for the options and attributes are chosen with our intuition of importance. For example, the implementation of \acp{POU} has more impact on their similarity than their variable declarations. 
In the following, we show all metrics employed in this evaluation in detail with their respective attributes and options.

\begin{figure}[ht!]
	\centering
	\includegraphics[width=\linewidth]{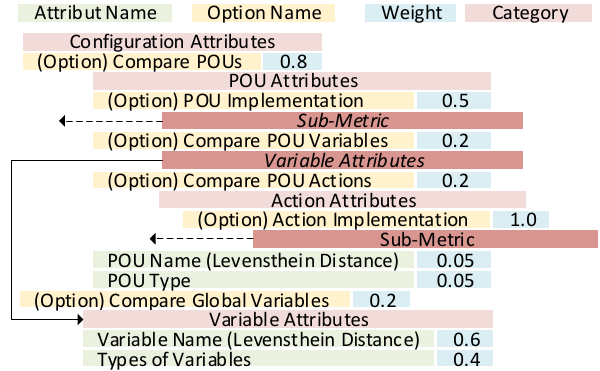}
	\caption{Illustration of the base metric with all selected options and attributes.}
	\label{tab:base_metric}
\end{figure}

\paragraph{\textbf{Base Metric}} Driven by the base metric, the comparison process decomposes models from \ac{POU} level down to fine-grained implementation artifacts such as statements.
The base metric is not used in the evaluation directly. Instead, it represents the common base for the following metrics that extend it.
\autoref{tab:base_metric} shows the base metric, which contains mostly options.
The black arrows on the left side indicate pointers to avoid redundancy and to support language nesting.
For example, the \emph{variable attributes} used in the \emph{POU variables} option are the same as the \emph{variable attributes} used in the \emph{global variables} option. Therefore, we simply place a pointer from the former to the latter.
The dashed, black arrows on the left side indicate pointers to options and attributes in sub-metrics.

\begin{figure}
	\centering
	\includegraphics[width=\linewidth]{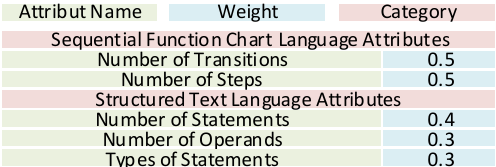}
	\caption{Illustration of the coarse-grained metric.}
	\label{tab:reve_metric}
\end{figure}

\paragraph{\textbf{Coarse-grained Metric}}
This metric is based on our previous work~\cite{schlie2019analyzing} and extends the base metric with attributes for the comparison of languages as illustrated in \autoref{tab:reve_metric}.
It is based on attributes that simply count elements, i.e., how many statements are contained in an \ac{ST} implementation or how many steps are performed in an \ac{SFC} implementation.

\begin{figure}[ht!]
	\centering
	\includegraphics[width=\linewidth]{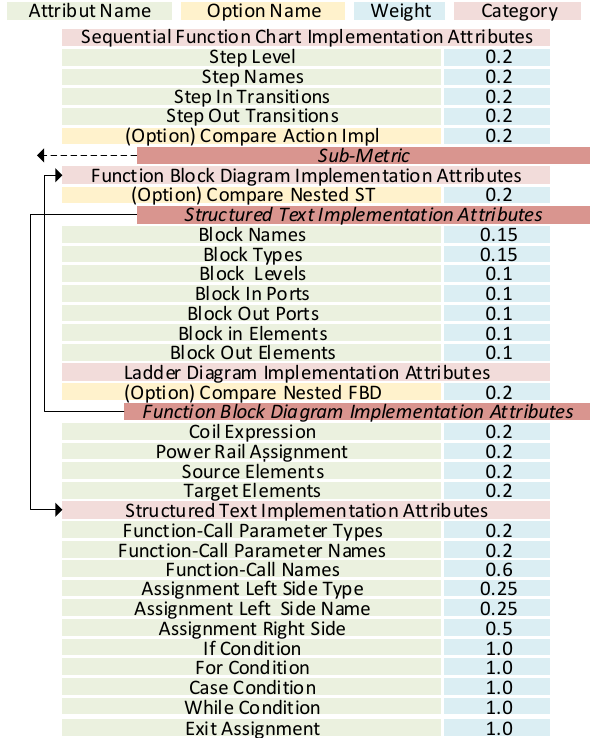}
	\caption{Illustration of the fine-grained metric.}
	\label{tab:ppu_metric}
\end{figure}

\paragraph{\textbf{Fine-grained Metric}}
The \emph{fine-grained metric} extends the base metric with further attributes and options for the comparison of IEC~61131-3 languages.
In contrast to the coarse-grained metric, it compares fine-grained artifacts with each other and can detect fine-grained changes such as additional expressions in a condition.
It also contains additional options for the comparison of nested artifacts.
\autoref{tab:ppu_metric} shows all selected options and weighted attributes.
Again, the black arrows on the left indicate pointers. For example, a \ac{LD} implementation can contain nested \ac{FBD}, as indicated by the \emph{Compare Nested FBD} option. Therefore, underneath the option \emph{Compare Nested FBD} we place a pointer to the \emph{\ac{FBD} Implementation} category of the metric. These, in turn, contain an option \emph{Compare Nested ST} with a pointer to the \emph{\ac{ST} Implementation} category of the metric.

\subsection{Subject Systems}
To evaluate our approach for the detection of intra variant as well as inter variant clone detection, we use the \acf{PPU} and \acf{xPPU} scenarios.
The \ac{PPU} handles and manipulates work pieces of different material.
It is a universal demonstrator for the study of evolution of \ac{aPS} \cite{PPUMaterial}.
It consists of 23 evolutionary steps, each referred to as a scenario.
All used \ac{PPU} and \ac{xPPU} scenarios are IEC~61131-3 projects exported in the PLCOpenXML format.
The scenarios were created using TwinCat3~\cite{TwinCAT}.

\begin{figure}[h]
	\centering
	\includegraphics[width=\linewidth]{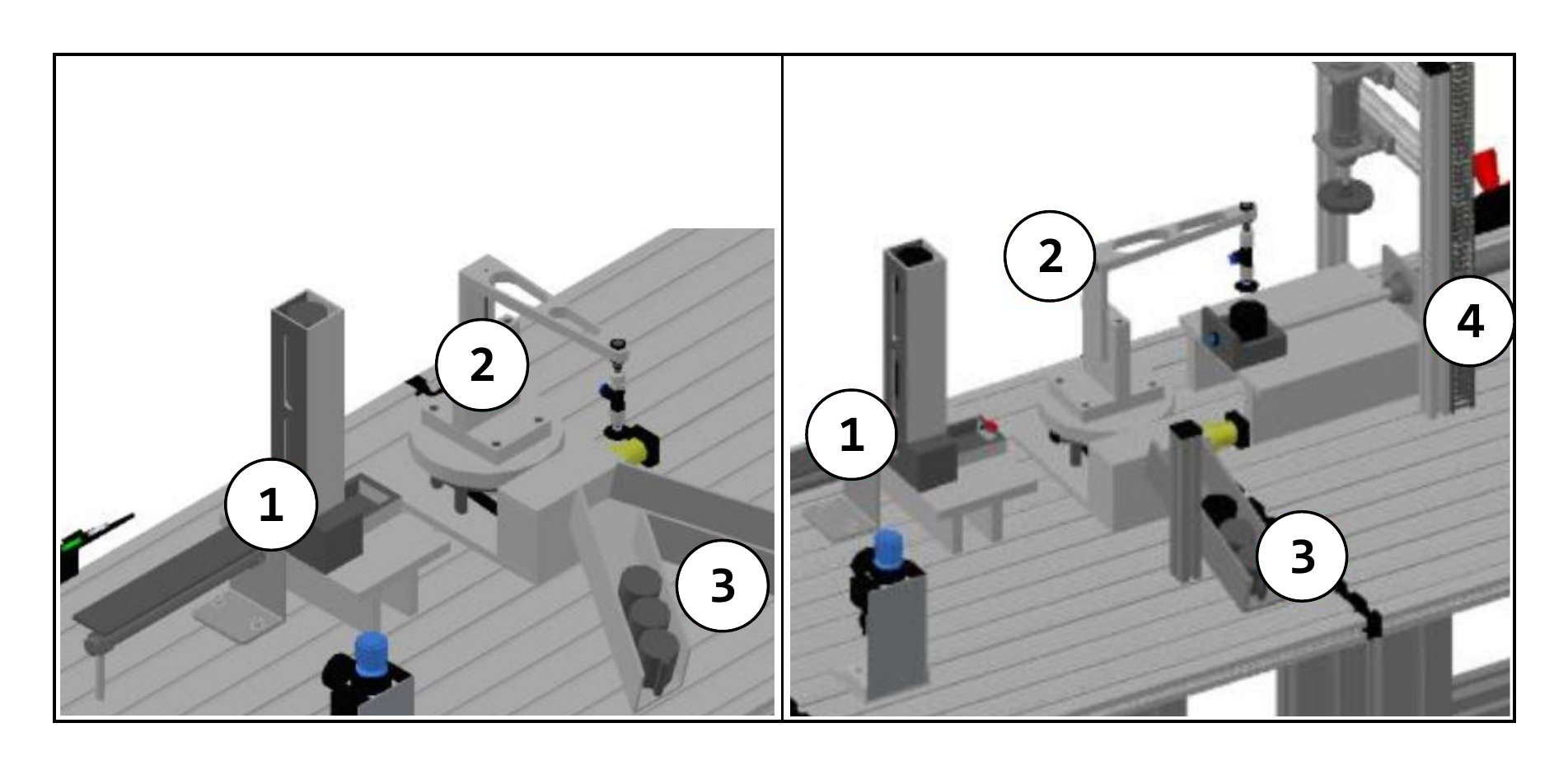}
	\caption{Illustration of the first PPU scenario on the left and the third PPU scenario on the right.}
	\label{fig:ppu_scenarios}
\end{figure}

\autoref{fig:ppu_scenarios} shows the first PPU scenario on the left and the third PPU scenario on the right.
Scenario 1 of the \ac{PPU} consists of a stack \circlearound{1} that serves as input storage for workpieces a crane \circlearound{2} utilized for the transportation of workpieces and a ramp \circlearound{3} as a workpiece output storage.
These parts can be found in the third scenario, as well. 
In the third scenario, a stamp \circlearound{4} was added to stamping workpieces.
The introduced functionality to stamp workpieces induces an adaption of the underlying software in response to the hardware changes.
For each scenario a \ac{PLC} implementation is available, which utilize \ac{ST} and \ac{SFC} as programming languages.
These scenarios contain language nesting between \ac{ST} and \ac{SFC}, which are actions that are called in \ac{SFC} and implemented in \ac{ST}.
Further information on the PPU and the respective evolution of the scenarios can be found in \cite{PPUMaterial}.

\subsection{Methodology}
The evaluation is divided into two parts: a quantitative and a qualitative analysis.
During the \emph{quantitative analysis}, we measure correctness and scalability.
To determine the \emph{correctness} (RQ1), we measure the precision and recall.
The measurement of recall and precision is difficult for tool developers due to a lack of case studies with an existing ground truth \cite{roy2014vision,roy2007survey}.
Therefore, we decided to employ a mutation-based strategy.
We implemented a mutation framework for IEC~61131-3 based programs.
The mutation framework is integrated into the \ac{VAT} and available on GitHub\footnote{\url{https://github.com/TUBS-ISF/variability\_analysis\_toolkit\_iec}}.

\begin{figure}[h]
	\centering
	\includegraphics[width=0.6\linewidth]{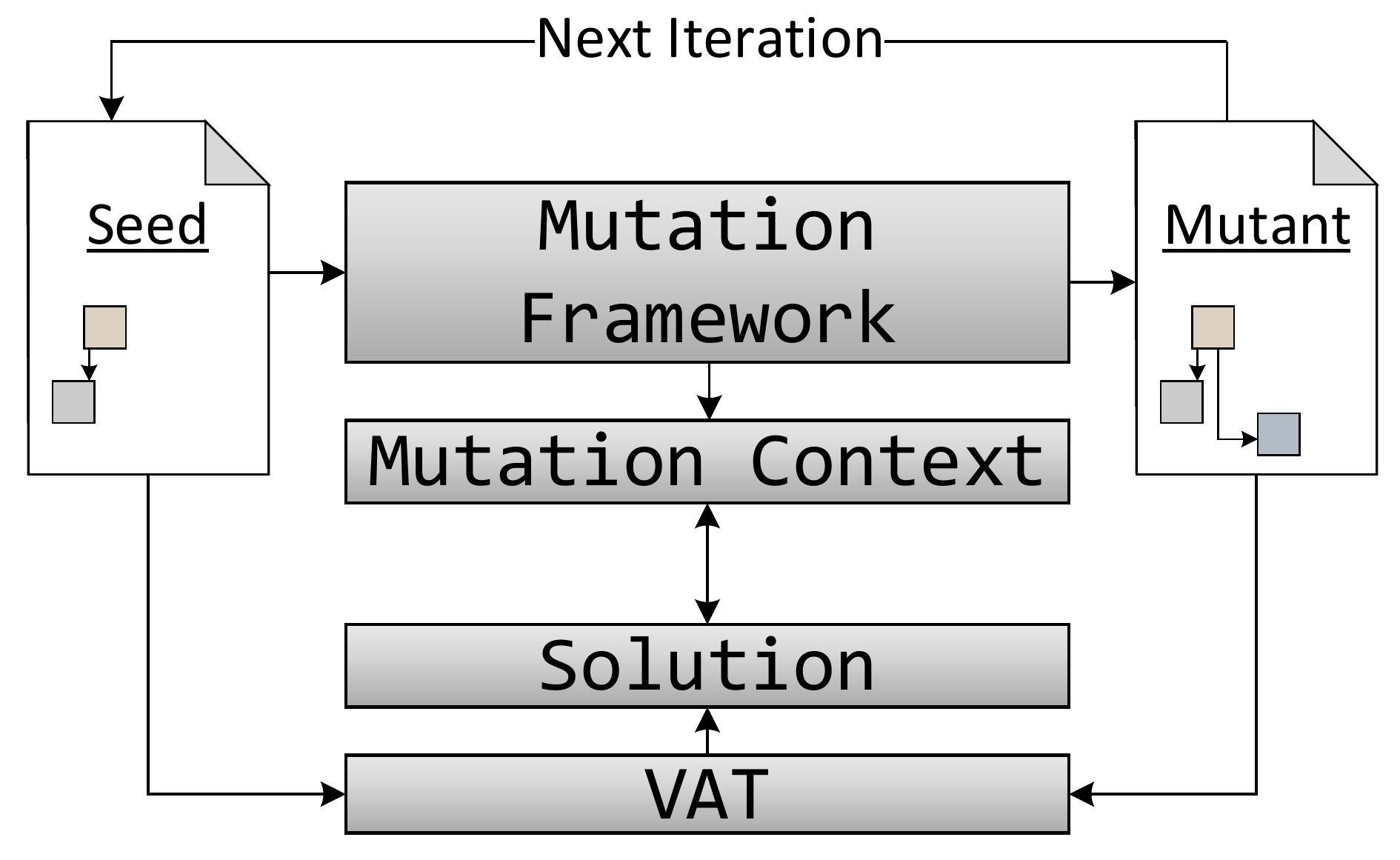}
	\caption{Iterative evaluation process to calculate the recall and precision of our clone detection approach.}
	\label{fig:mutation_framework}
\end{figure}

The evaluation is driven by a evaluation-cycle shown in \autoref{fig:mutation_framework}.
The process starts with a randomly selected model of the \ac{PPU} and \ac{xPPU} scenarios, which serves as a seed for the mutation process.
Our mutation framework creates an exact copy of this seed model and randomly injects mutations.
We defined 11 mutation-operators for the mutation to add or remove artifacts and change identifiers or expressions.
All mutations are based on the meta-model representation that reflects the IEC~61131-3 standard and generates syntactically correct artifacts.
Changes are stored as pairs in the mutation context: $mutation = (origin, mutant, operator)$.
Mutated scenarios can be stored as models and the mutation context as JSON files, allowing to generate cloned variants with existing ground truth.
To evaluate our approach, we devised an automated process, which generates mutants, compares them, and calculates the precision and recall.
The process generates a mutant out of a selected seed model and stores all changed artifacts in a mutation context.
Seed model and mutant model are compared using the \ac{VAT}.
Changed elements are collected out of the resulting solution data-structure and compared with the mutation context.
The precision and recall are measured for only the injected/changed artifacts.
We use the following definition for the interpretation of our results:
Changed artifacts within the mutation context and solution data-structure are true positives (TP).
Changed artifacts in the solution data-structure, but not in the mutation context are false positives (FP).
Artifacts that are only in the mutation context and not in the solution data-structure are false negatives (FN).
For the calculation of the precision and recall, we use the following definitions:
\begin{equation*}
\text{Precision} = \frac{TP}{TP + FP} \quad\quad \text{Recall} = \frac{TP}{TP+FN}
\end{equation*}
%
To evaluate the \emph{scalability} (RQ2) of the \ac{VAT}, we measure run time and memory consumption during the comparison approach in relation to the system size. 

The second part of our evaluation is the \emph{qualitative analysis}, which gives insight into the \emph{usefulness} (RQ3) of the \ac{VAT}.
We perform a pairwise comparison of each \ac{PPU} and \ac{xPPU} scenario using the coarse-grained as well as the fine-grained metric.
To evaluate our approach for intra-system clone detection, we perform a clone analysis on the granularity level of \acp{POU} for each \ac{PPU} and \ac{xPPU} variant.
The intra-system clone analysis granularity is adjustable, for example to \ac{POU} level or even finer-grained implementation level.
For the purpose of this evaluation, we decided to visualize the results on the granularity level of \acp{POU}, as it is well-suited for a re-engineering workflow towards planned reuse of IEC~61131-3 legacy software~\cite{fischer2020reengineering}.

\subsection{Results and Discussion}
In this section, we present and discuss the results of our quantitative and qualitative evaluation.

\subsubsection{Quantitative Analysis: Correctness (RQ1)}

To evaluate correctness, we use our mutation framework to fully automatically compute recall and precision using a standard mutation-based analysis procedure.
As per standard for mutation analysis, the framework is configured to only use one category of mutations, i.e., either only T2 mutations such as renaming of artifacts, or only T3 mutations such as insertions or deletions of artifacts, per model.
We perform two runs with 10000 iterations each, to determine the precision and recall for Type~II and Type~III mutations respectively.
We also calculate the overall precision an recall over all types of mutations.
Due to the parsing process which normalizes the PLCOpenXML files, we cannot detect Type~I mutations, and thus not distinguish Type~I and Type~II clones.
\begin{table}[h]
	\centering
	\resizebox{\linewidth}{!}{
	\begin{tabular}{c||cc|cc|cc}
		\toprule
		\textbf{}& \textbf{Fine T2}& \textbf{Coarse T2} & \textbf{Fine T3}& \textbf{Coarse T3}& \textbf{Fine Total}& \textbf{Coarse Total} \\
		\midrule
		\midrule
		\textbf{True Positives}     & 7738		& 686		  & 9991	&6852	&17729	&7538      \\
		\textbf{False Positives}    & 0			& 0      	  & 0  		&44		&0		&44      \\
		\textbf{False Negatives}    & 2262 		& 9314		  & 9	  	&3148	&2271	&12462      \\
		\midrule
		\textbf{Precision-Total}    & 100 		& 100	  	 & 100	  	&99,36	&\textbf{100}	&\textbf{99,42}      \\
		\textbf{Recall-Total}		& 77,38		& 6,86       & 99,91	&68,52	&\textbf{88,65}	&\textbf{37,69}      \\
		\midrule	
	\end{tabular}
	}
	\caption{Evaluated precision and recall for the fine- and coarse-grained metric with all resulting value based on 20.000 mutations.}
	\label{tab:precision_and_recall}
\end{table}
In \autoref{tab:precision_and_recall}, we show the precision and recall measures.

The coarse-grained metric achieves a precision of 100\% for the analysis of Type~II clones but only a recall of 6,86\%.
Only coarse changes on configuration or \ac{POU} level could be detected.
Renaming of implementation artifacts could not be detected due to the fact that elements are only counted and not compared directly.
For the analysis of Type III clones, the coarse metric achieves a higher recall of 68,52\%, as added or removed artifacts result in a changed similarity value during the comparison between those elements and can thus be detected.
Overall, the coarse-grained metric achieves a precision of 99,42\% and a recall of 37,69\%.

The fine-grained metric achieves better results, as expected.
For the analysis of Type II clones, precision is at 100\% and recall at 77,38\%.
A detailed analysis of the resulting mutants showed that the \ac{VAT}, using the fine-grained metric for the comparison, had problems with the renaming of some elements such as the returning type of a \ac{POU} within the Type II run.
To improve the recall value in this case, additional attributes are needed for the comparison of \acp{POU} that consider those elements.
For the detection of Type III clones, the \ac{VAT} achieved a precision of 100 \% and a recall of 99,91\%.
Some variable insertions and removals could not be detected, causing all 9 false negatives.
Overall, our approach achieved a precision of 100\% and a recall of 88,65\% using the fine-grained metric, which are excellent precision and recall values.

The fine-grained metric could, in contrast to the coarse-grained metric, detect changes on the implementation level, such as renamed, added, or removed statements.
We conclude that, based on a coarse-grained metric, IEC~61131-3 systems can be analyzed only on an abstraction level where the correct position of cloned artifacts is not essential, which is why the coarse-grained metric has such a low recall value.
The fine-grained metric is the better choice to get more details about the correct location of code clones.
Ultimately, the desired granularity of comparison can be freely configured based on the comparison metric to achieve the desired results.
Detected clones are then shown in the family model at a desired level of granularity.

Overall, the used data set together with the achieved results provide strong evidence for the correctness of our approach.
Moreover, the created metrics can be automatically evaluated using the mutation framework to assess their feasibility.

\subsubsection{Quantitative Analysis: Scalability (RQ2)}

\paragraph{\textbf{Run Time (RQ 2.1)}}
Reasonable run-times for the analysis of fine-grained variability relations between software variants are required for the productive use of the \ac{VAT}. 
To measure the run-time, we use an event-based benchmark system that is already implemented in the \ac{VAT}.
For the comparison, we created a metric that compares all artifacts within the \ac{xPPU} scenarios.

\begin{figure}[ht!]
	\centering
	\includegraphics[width=\linewidth]{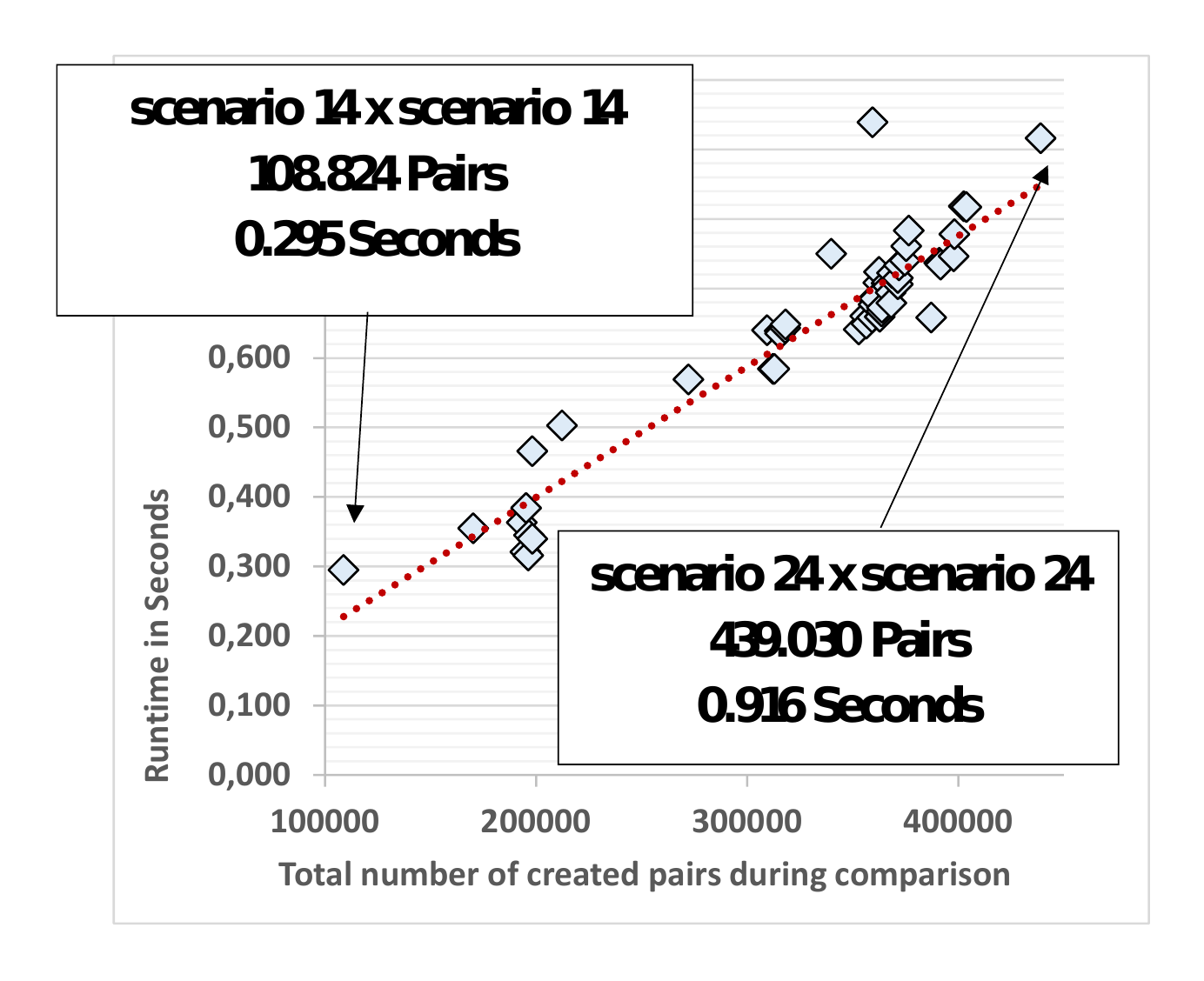}
	\caption{Run-time in relation to number of created pairs during the comparison.}
	\label{fig:run_time_pairwise}
\end{figure}

We perform all possible pairwise comparisons of the \ac{xPPU} scenarios and show the run-time results in \autoref{fig:run_time_pairwise}.
The x-axis shows the total number of element pairs that were created during the comparison process, and the y-axis shows the required time for the comparison in seconds.
On the left side in \autoref{fig:run_time_pairwise}, we can see the comparison of scenario S14 with itself.
Scenario S14 is the smallest scenario of the \ac{xPPU} scenarios, and the comparison process creates $\approx 108,000$ element pairs when compared with itself, which takes $0.295$ seconds.
The comparison of the largest \ac{xPPU} scenario S24 with itself is shown on the right in \autoref{fig:run_time_pairwise} and in detail in \autoref{fig:run_time_pairwise}.
During this comparison, $\approx 439,000$ element pairs are created, and it requires $0.916$ seconds.
The red line indicates a linear relationship between created element pairs and comparison time.

\begin{table}
	\centering
		\resizebox{0.8\linewidth}{!}{
	\begin{tabular}{cccc}
		\toprule
		\textbf{Elements}& \textbf{Pairs}& \textbf{Attributes} & \textbf{Comparisons}\\
		\midrule
		\midrule
		\textbf{Projects}   & 1 	        & 0     & 0	        \\
		\textbf{POU} 		& 2,116         & 2     & 4,232         \\
		\textbf{Actions} 	& 25,663        & 0     & 0         \\
		\textbf{Steps}		& 207,125       & 4 	& 828,500   \\
		\textbf{Statements} & 204,126       & 10  	& 2,041,260 \\
		\midrule
		\midrule
		\textbf{Overall}   & 439,030  &	& 2,873,992   \\
		\midrule	
	\end{tabular}
	}
	\caption{Pairs of created elements and comparisons for the detection of clones within the largest xPPU \emph{scenario S24}.}
	\label{tab:results_pairs}
\end{table}

To provide more details about the comparison, we count the number of element pairs that are created during the comparison of \emph{scenario S24} with itself. 
This scenario represents the last evolution step of the \ac{xPPU} scenarios and contains all elements of the scenarios before.
In \autoref{tab:results_pairs}, we list the number of element pairs that are created and the number of attributes and comparisons.
The comparison process creates $\approx 207,000$ pairs of steps.
These steps are compared with four attributes, which results in over $828,500$ comparisons. 
Overall, for the comparison of \emph{scenario S24} with itself, the compare process creates $\approx 439,030$ pairs of elements.
In total, to compare both models, $\approx 2.9$ million comparisons are made.
To reduce the random impact of a non-closed test system, we repeated the run-time measurement ten times and created average values.
For the comparison of \emph{scenario S24} with itself, the comparison process takes a total time of $0.916$ seconds on average. 
The trend-line in \autoref{fig:run_time_pairwise} indicates a linear increase in run-time in relation to artifacts to compare.
Hence, we consider the run-time of the \ac{VAT} as reasonable for the evaluated scenarios. 

\paragraph{\textbf{Memory Consumption (RQ 2.2)}}
Another key factor for the scalability of software, such as the \ac{VAT}, is the memory consumption.
We assume that, if the \ac{VAT} is reliably applicable to realistic industrial models, then it is useful to domain experts.

\begin{figure}[h]
	\centering
	\includegraphics[width=\linewidth]{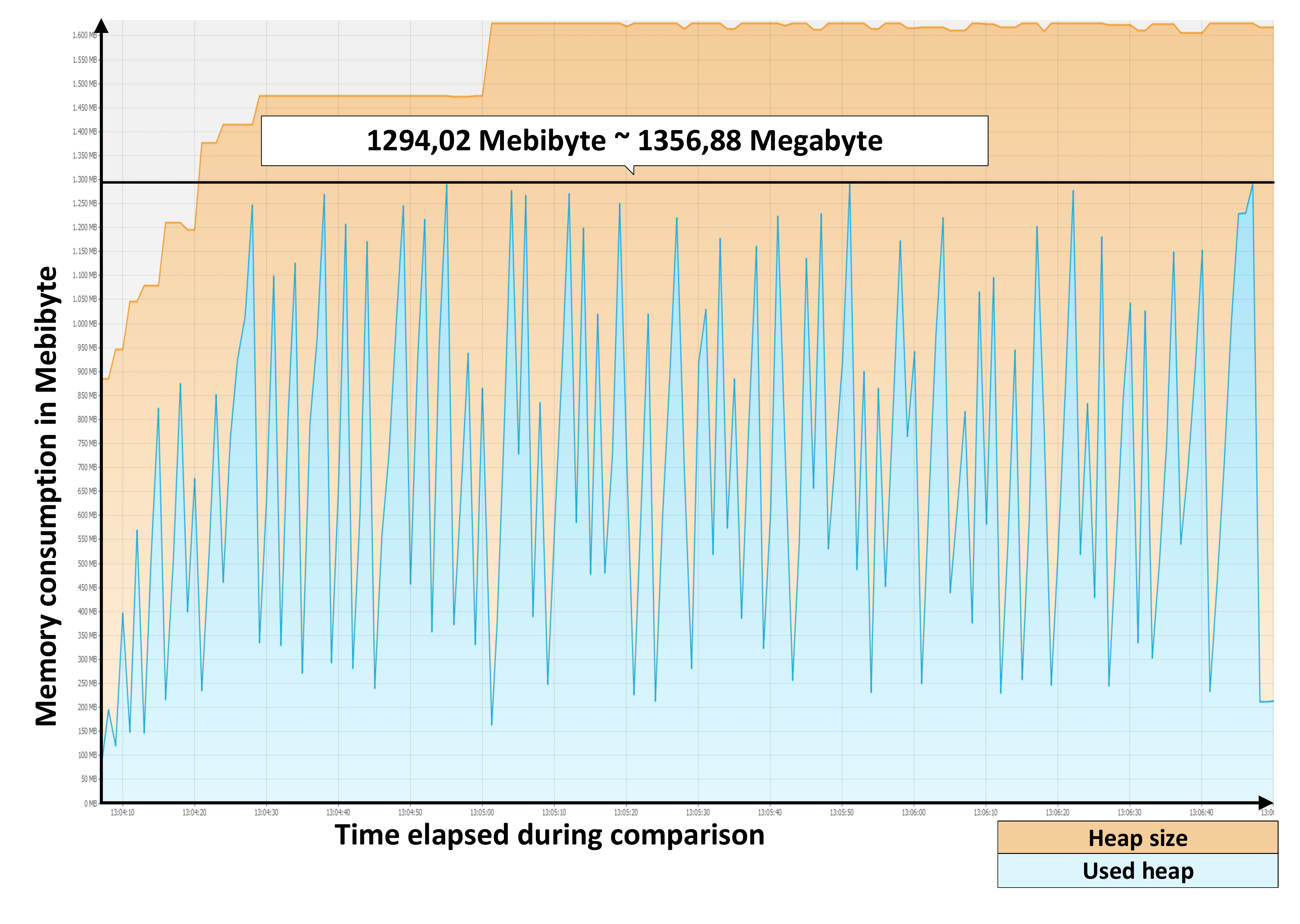}
	\caption{Memory consumption (y axis) of the \ac{VAT} when applied to all PPU and xPPU scenarios in sequence (x axis).}
	\label{fig:heap_dump}
\end{figure}

To determine the memory consumption, we use \emph{VisualVM} that provides insight into the \ac{JVM} memory dump assigned to a process ID.
We perform a scripted pairwise comparison of all \ac{PPU} scenarios.
Specifically, we compare each of the 23 \ac{PPU} scenarios pairwise, resulting in a total of 276 pairwise comparisons.
This means we perform the comparison and matching steps and storage of the family models 276 times.
The memory consumption is shown in \autoref{fig:heap_dump}.
The orange area represents the allocated heap memory of the \ac{JVM}.
The blue area represents the used heap, i.e., how much memory is actually used by the \ac{VAT}.
For all 276 comparisons, the process takes 3 minutes and 18 seconds.
The initial heap size is $\approx 850$ MB, and during the comparison, the \ac{JVM} allocated more memory to a maximum of $\approx 1,697$ MB.
After each comparison of two models, the family model is created and drawn to the family model view.
Each peak in used memory is a result of the drawing process.
The maximum amount of memory used depends on the size of the drawn models.
Comparison of \emph{scenario S1} with itself results in a smaller family model than the comparison between \emph{scenario S24} with itself. Consequently, the memory consumption is lower.
However, as a maximum value, $\approx 1,356$ MB of RAM is used for the pairwise comparison of 23 models.
Thus, we conclude that our approach scales for larger models as well.

\subsubsection{Qualitative Analysis: Usefulness (RQ3)}
\label{sec:qualitative}
Based on the quantitative analysis we could show that the output that is created by the \ac{VAT} is correct and that the \ac{VAT} scales well.
In this section, we show a qualitative analysis to assess the usefulness of the \ac{VAT} for the re-engineering of IEC~61131-3 based systems by giving insights into the detected clones in the PPU and xPPU evolution scenarios.

\paragraph{\textbf{Scenario Similarity (RQ 3.1)}}
We separately evaluate the overall similarities calculated for each pairwise project comparison for the PPU and xPPU scenario sets. 
The overall similarity is an indicator of the relationship between projects.
We created and analyzed 292 family models for the analysis of PPU and xPPU scenarios in total.
In this section, we only provide aggregated data and refer to our supplementary material\footnote{\url{https://github.com/TUBS-ISF/IEC\_61131\_3\_Clone\_Detection}}, which contains all scenarios, metrics, and family models, and the implementation of the VAT.

\begin{table}
	\centering
	\resizebox{\linewidth}{!}{
		\begin{tabular}{c|cccccccccccc}
			\hhline{~|*{12}{-}}
			\textbf{\%}  & \textbf{S2} & \textbf{S3} & \textbf{S4a} & \textbf{S4b} & \textbf{S5} & \textbf{S7} & \textbf{S8} & \textbf{S9} & \textbf{S10} & \textbf{S11} & \textbf{S12} & \textbf{S13}\\
			\hhline{~|*{12}{-}}
			\textbf{S1}  &91,21&75,05&71,52&71,51&75,08&65,97&65,97&63,1&63,1&63,1&63,1&60,32\\
			
			\textbf{S2}  & \diagbox{}{} &81,31&77,41&77,39&81,34&71,11&71,11&67,83&67,83&67,83&67,83&64,79\\
			
			\textbf{S3} & \diagbox{}{} & \diagbox{}{}&94,86&94,88&99,96&86,19&86,06&81,50&81,5&81,50&81,57&77,60\\
			
			\textbf{S4a} & \diagbox{}{} & \diagbox{}{} & \diagbox{}{} &99,58&94,82&89,46&89,33&84,65&84,65&84,65& 84,65&80,90\\
			
			\textbf{S4b}  & \diagbox{}{} & \diagbox{}{} & \diagbox{}{} & \diagbox{}{} &94,84&89,48&89,35&84,67&84,67&84,67&84,67&80,89\\
			
			\textbf{S5}  & \diagbox{}{} & \diagbox{}{} & \diagbox{}{} & \diagbox{}{} & \diagbox{}{} &86,16&86,03&81,48&81,47&81,47&81,47&77,58\\
			
			\textbf{S7}  & \diagbox{}{} & \diagbox{}{} & \diagbox{}{} & \diagbox{}{} & \diagbox{}{} & \diagbox{}{} &99,86&94,48&94,48&94,48&94,48&88,41 \\
			
			\textbf{S8}  & \diagbox{}{} & \diagbox{}{} & \diagbox{}{} & \diagbox{}{} & \diagbox{}{} & \diagbox{}{} & \diagbox{}{} &94,55&94,55&94,55&94,55&88,48 \\
			
			\textbf{S9}  & \diagbox{}{} & \diagbox{}{} & \diagbox{}{} & \diagbox{}{} & \diagbox{}{} & \diagbox{}{} & \diagbox{}{} & \diagbox{}{} &99,07&99,08&99,08&92,66 \\
			
			\textbf{S10}  & \diagbox{}{} & \diagbox{}{} & \diagbox{}{} & \diagbox{}{} & \diagbox{}{} & \diagbox{}{} & \diagbox{}{} & \diagbox{}{} & \diagbox{}{} &99,86&99,78&93,33 \\
			
			\textbf{S11}  & \diagbox{}{} & \diagbox{}{} & \diagbox{}{} & \diagbox{}{} & \diagbox{}{} & \diagbox{}{} & \diagbox{}{} & \diagbox{}{} & \diagbox{}{} & \diagbox{}{} &99,91&93,45\\
			
			\textbf{S12} & \diagbox{}{} & \diagbox{}{} & \diagbox{}{} & \diagbox{}{} & \diagbox{}{} & \diagbox{}{} & \diagbox{}{} & \diagbox{}{} & \diagbox{}{} & \diagbox{}{} & \diagbox{}{} &93,54 \\
			\hline
		\end{tabular}
	}
	\caption{Similarities between \emph{PPU} scenarios in percent using the fine-grained metric.}
	\label{tab:ppu_similarity}
\end{table}	

\begin{table}
	\centering
	\resizebox{\linewidth}{!}{
		\begin{tabular}{c|cccccccccc}
			\hhline{~|*{10}{-}}
			\textbf{\%}  & \textbf{S15} & \textbf{S16} & \textbf{S17} & \textbf{S18} & \textbf{S19} & \textbf{S20} & \textbf{S21}& \textbf{S23}  & \textbf{S24}\\
			\hhline{~|*{10}{-}}	
			\textbf{S14} 	&78,33&72,40&68,90&67,58&65,43&55,45&52,17&51,46&51,09\\
			\textbf{S15}  & \diagbox{}{} &91,43&86,54&84,82&81,15&67,8&63,61&62,64&62,22\\
			\textbf{S16}  & \diagbox{}{} & \diagbox{}{} &94,51&92,57&88,22&73,34&68,63&67,56&67,14 \\
			\textbf{S17}  & \diagbox{}{} & \diagbox{}{} &\diagbox{}{}&97,18&91,73&76,07&71,18&70,05&69,63\\
			\textbf{S18}  & \diagbox{}{} & \diagbox{}{} &\diagbox{}{}&\diagbox{}{}&93,87&77,75&72,74&71,58&71,16 \\
			\textbf{S19}  & \diagbox{}{} & \diagbox{}{} &\diagbox{}{}&\diagbox{}{}&\diagbox{}{} &82,56&77,20&75,94&75,40 \\	
			\textbf{S20}  & \diagbox{}{} & \diagbox{}{} & \diagbox{}{} & \diagbox{}{} & \diagbox{}{} & \diagbox{}{} &93,64&92,02&91,49 \\
			\textbf{S21} & \diagbox{}{} & \diagbox{}{} & \diagbox{}{} & \diagbox{}{} & \diagbox{}{} & \diagbox{}{}  & \diagbox{}{} &98,20&96,46 \\
			\textbf{S23}  & \diagbox{}{} & \diagbox{}{} & \diagbox{}{} & \diagbox{}{} & \diagbox{}{} & \diagbox{}{} & \diagbox{}{} & \diagbox{}{} & 96,70 \\
			
			\hline
		\end{tabular}
	}
	\caption{Similarities between \emph{xPPU} scenarios in percent using the fine-grained metric.}
	\label{tab:xppu_similarity}
\end{table}

In \autoref{tab:ppu_similarity} and \autoref{tab:xppu_similarity}, we show all pairwise similarities for the comparison of the PPU and xPPU scenarios.
In both cases, the similarity decreases in every row from left to right, which is plausible and can be explained by the increasing divergence of scenarios that are further apart in time.
\begin{figure}[h]
	\centering
	\includegraphics[width=0.8\linewidth]{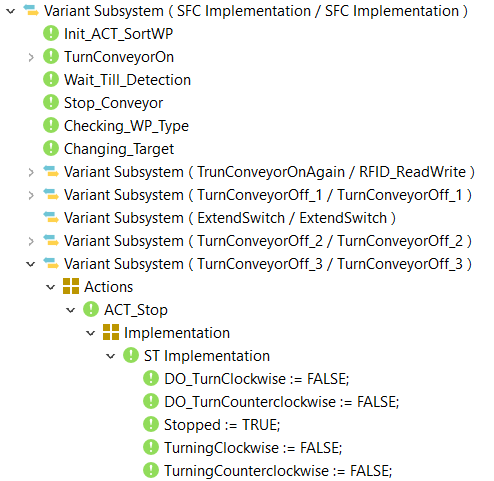}
	\caption{Snippet of the family model of the comparison between scenarios S23 and S24.}
	\label{fig:findings_1}
\end{figure}
In \autoref{fig:findings_1}, we show a fragment of the resulting family model of comparing scenarios S23 and S24.
The \emph{TurnConveyorOff\_3} step, which is used by the \ac{SFC} implementation, calls an action that is implemented in \ac{ST}.
The resulting family model shows that the nested action could be matched, which results in a mandatory element.
Moreover, we could analyze different scenarios that show different kinds of language nesting, such as a utilized function block in a \ac{LD} implementation or a \ac{ST} implementation in a function block.
Hence, we argue that our comparison approach can detect and compare nested implementations and that it is useful in real-world scenarios.
\begin{figure}[h]
	\centering
	\includegraphics[width=\linewidth]{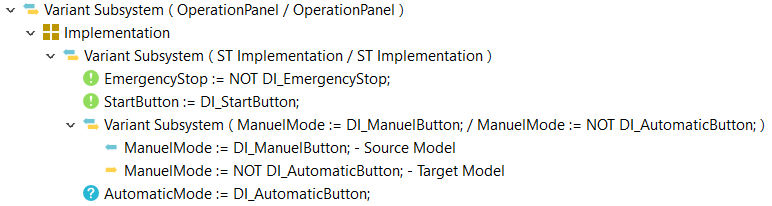}
	\caption{Snippet of the family model of the comparison between scenarios S14 and S23.}
	\label{fig:findings_2}
\end{figure}
Overall, we were able to detect fine-grained changes such as parts of a statement like renaming of a function or an additional literal in an expression.
In \autoref{fig:findings_2}, we show a snippet that shows the resulting family model of the comparison between scenarios S14 and S23.
We can see a variation point between both \emph{OperationPanel} \acp{POU}, which contains two mandatory \mandatory{}, a changed \alternative{}, and an additional \optional{} assignment.
We investigated this change and found that it can be explained with a renaming of the \emph{DI\_ManuelButton} into \emph{DI\_AutomaticButton}.
We analyzed all resulting family models and the respective PLCOpenXML files to manually verify our results and found them to be sensible.

\begin{table}
	\centering
	\resizebox{\linewidth}{!}{
		\begin{tabular}{c|cccccccccccc}
			\hhline{~|*{12}{-}}
			\textbf{\%}  & \textbf{S2} & \textbf{S3} & \textbf{S4a} & \textbf{S4b} & \textbf{S5} & \textbf{S7} & \textbf{S8} & \textbf{S9} & \textbf{S10} & \textbf{S11} & \textbf{S12} & \textbf{S13}\\
			\hhline{~|*{12}{-}}
			\textbf{S1}  &\cellcolor{red!40}-1,07&\cellcolor{red!80}-1,92&\cellcolor{red!40}-1,82&\cellcolor{red!40}-1,81&\cellcolor{red!80}-1,92&\cellcolor{red!40}-1,77&\cellcolor{red!40}-1,77&\cellcolor{red!40}-1,85&\cellcolor{red!40}-1,85&\cellcolor{red!40}-1,85&\cellcolor{red!40}-1,85&\cellcolor{red!40}-1,43\\
			
			\textbf{S2}  & \diagbox{}{} &\cellcolor{red!40}-1,45&\cellcolor{red!40}-1,38&\cellcolor{red!40}-1,38&\cellcolor{red!40}-1,45&\cellcolor{red!40}-1,44&\cellcolor{red!40}-1,44&\cellcolor{red!40}-1,61&\cellcolor{red!40}-1,61&\cellcolor{red!40}-1,61&\cellcolor{red!40}-1,61&\cellcolor{red!40}-1,43\\
			
			\textbf{S3}  & \diagbox{}{} & \diagbox{}{}&-0,02\cellcolor{red!20}&\cellcolor{red!20}-0,01&\cellcolor{blue!20}0,00&\cellcolor{red!20}-0,57&\cellcolor{red!20}-0,57&\cellcolor{red!20}-1,00&\cellcolor{red!20}-1,00&\cellcolor{red!20}-1,00&\cellcolor{red!20}-1,00&\cellcolor{red!20}-0,61\\
			
			\textbf{S4a}  & \diagbox{}{} & \diagbox{}{} & \diagbox{}{} &\cellcolor{red!20}-0,02&\cellcolor{red!20}-0,02&\cellcolor{green!20}0,32&\cellcolor{green!20}0,32&\cellcolor{red!20}-0,15&\cellcolor{red!20}-0,15&\cellcolor{red!20}-0,15&\cellcolor{red!20}-0,15&\cellcolor{green!20}0,21 \\
			
			\textbf{S4b}  & \diagbox{}{} & \diagbox{}{} & \diagbox{}{} & \diagbox{}{} &\cellcolor{red!20}-0,01&\cellcolor{green!80}0,33&\cellcolor{green!80}0,33&\cellcolor{red!20}-0,14&\cellcolor{red!20}-0,14&\cellcolor{red!20}-0,14&\cellcolor{red!20}-0,14&\cellcolor{green!20}0,29\\
			
			\textbf{S5}  & \diagbox{}{} & \diagbox{}{} & \diagbox{}{} & \diagbox{}{} & \diagbox{}{} &\cellcolor{red!20}-0,57&\cellcolor{red!20}-0,57&\cellcolor{red!20}-1,00&\cellcolor{red!40}-1,01&\cellcolor{red!40}-1,01&\cellcolor{red!40}-1,01&\cellcolor{red!20}-0,6\\
			
			\textbf{S7}  & \diagbox{}{} & \diagbox{}{} & \diagbox{}{} & \diagbox{}{} & \diagbox{}{} & \diagbox{}{} &\cellcolor{blue!20}0,00&\cellcolor{red!20}-0,54&\cellcolor{red!20}-0,54&\cellcolor{red!20}-0,54&\cellcolor{red!20}-0,54&\cellcolor{red!40}-1,7 \\
			
			\textbf{S8}  & \diagbox{}{} & \diagbox{}{} & \diagbox{}{} & \diagbox{}{} & \diagbox{}{} & \diagbox{}{} & \diagbox{}{} &\cellcolor{red!20}-0,54&\cellcolor{red!20}-0,54&\cellcolor{red!20}-0,54&\cellcolor{red!20}-0,54&\cellcolor{red!40}-1,7 \\
			
			\textbf{S9}  & \diagbox{}{} & \diagbox{}{} & \diagbox{}{} & \diagbox{}{} & \diagbox{}{} & \diagbox{}{} & \diagbox{}{} & \diagbox{}{} &\cellcolor{red!20}-0,01&\cellcolor{red!20}-0,01&\cellcolor{red!20}-0,01&\cellcolor{red!40}-1,20 \\
			
			\textbf{S10} & \diagbox{}{} & \diagbox{}{} & \diagbox{}{} & \diagbox{}{} & \diagbox{}{} & \diagbox{}{} & \diagbox{}{} & \diagbox{}{} & \diagbox{}{} &\cellcolor{red!20}-0,01&\cellcolor{red!20}-0,02&\cellcolor{red!40}-1,21 \\
			
			\textbf{S11}  & \diagbox{}{} & \diagbox{}{} & \diagbox{}{} & \diagbox{}{} & \diagbox{}{} & \diagbox{}{} & \diagbox{}{} & \diagbox{}{} & \diagbox{}{} & \diagbox{}{} &\cellcolor{red!20}-0,01&\cellcolor{red!40}-1,20\\
			
			\textbf{S12}  & \diagbox{}{} & \diagbox{}{} & \diagbox{}{} & \diagbox{}{} & \diagbox{}{} & \diagbox{}{} & \diagbox{}{} & \diagbox{}{} & \diagbox{}{} & \diagbox{}{} & \diagbox{}{} & \cellcolor{red!40}-1,19\\
			\hline
		\end{tabular}
	}
	\resizebox{\linewidth}{!}{
		\begin{tabular}{ccccccccccccccc}
			\multicolumn{3}{c}{lowest similarity} &min\cellcolor{red!80}&-3\cellcolor{red!60} &\cellcolor{red!40}-2&\cellcolor{red!20}-1&\cellcolor{blue!20}0&\cellcolor{green!20}1 &\cellcolor{green!40}2&\cellcolor{green!60}3&\cellcolor{green!80}max&\multicolumn{3}{c}{highest similarity}\\
		\end{tabular}	
	}
	
	\caption{Similarity difference heat-map of PPU scenarios: similarity is higher than with the coarse-grained metric marked green and lower marked red.}
	\label{tab:ppu_similarity_diff}
\end{table}	

\begin{table}
	\centering
	\resizebox{\linewidth}{!}{
		\begin{tabular}{c|ccccccccc}
			\hhline{~|*{9}{-}}
			\textbf{\%}  & \textbf{S15} & \textbf{S16} & \textbf{S17} & \textbf{S18} & \textbf{S19} & \textbf{S20} & \textbf{S21} & \textbf{S23} & \textbf{S24}\\
			\hhline{~|*{9}{-}}	
			\textbf{S14}  & \cellcolor{red!40}-1,43&\cellcolor{red!40}-1,32&\cellcolor{red!40}-1,43&\cellcolor{red!40}-1,2&\cellcolor{red!60}-2,25&\cellcolor{red!40}-1,62&\cellcolor{red!40}-1,51&\cellcolor{red!40}-1,48&\cellcolor{red!40}-1,39\\
			\textbf{S15}  & \diagbox{}{} &\cellcolor{red!20}-0,21&\cellcolor{red!20}-0,53&\cellcolor{red!20}-0,19&\cellcolor{red!40}-1,91&\cellcolor{red!40}-1,36&\cellcolor{red!40}-1,26&\cellcolor{red!40}-1,24&\cellcolor{red!40}-1,17\\
			\textbf{S16}  & \diagbox{}{} & \diagbox{}{} &\cellcolor{red!20}-0,38&\cellcolor{red!20}-0,01&\cellcolor{red!60}-2,08&\cellcolor{red!40}-1,5&\cellcolor{red!40}-1,39&\cellcolor{red!40}-1,36&\cellcolor{red!40}-1,29 \\
			\textbf{S17}  & \diagbox{}{} & \diagbox{}{} &\diagbox{}{}&\cellcolor{red!20}-0,36&\cellcolor{red!80}-3,00&\cellcolor{red!60}-2,22&\cellcolor{red!60}-2,07&\cellcolor{red!60}-2,02&\cellcolor{red!40}-1,95\\
			\textbf{S18}  & \diagbox{}{} & \diagbox{}{} &\diagbox{}{}&\diagbox{}{}&\cellcolor{red!60}-2,95&\cellcolor{red!60}-2,18&\cellcolor{red!60}-2,03&\cellcolor{red!40}-1,98 &\cellcolor{red!40}-1,91\\
			\textbf{S19}  & \diagbox{}{} & \diagbox{}{} &\diagbox{}{}&\diagbox{}{}&\diagbox{}{} &\cellcolor{green!20}0,14&\cellcolor{green!20}0,14&\cellcolor{green!20}0,13&\cellcolor{green!80}0,22 \\
			
			\textbf{S20}  & \diagbox{}{} & \diagbox{}{} & \diagbox{}{} & \diagbox{}{} & \diagbox{}{}  & \diagbox{}{} &\cellcolor{blue!20}0,00&\cellcolor{blue!20}0,00&\cellcolor{green!20}0,09 \\
			
			\textbf{S21}  & \diagbox{}{} & \diagbox{}{} & \diagbox{}{} & \diagbox{}{} & \diagbox{}{} & \diagbox{}{} & \diagbox{}{} &\cellcolor{red!20}-0,04 &\cellcolor{red!20}-0,81 \\
			\textbf{S23}  & \diagbox{}{} & \diagbox{}{} & \diagbox{}{} & \diagbox{}{} & \diagbox{}{} & \diagbox{}{} & \diagbox{}{} &\diagbox{}{} &\cellcolor{red!40}-1,51  \\
			\hline
		\end{tabular}	
	}
	\resizebox{\linewidth}{!}{
		\begin{tabular}{ccccccccccccccc}
			\multicolumn{3}{c}{lowest similarity} &min\cellcolor{red!80}&-3\cellcolor{red!60} &\cellcolor{red!40}-2&\cellcolor{red!20}-1&\cellcolor{blue!20}0&\cellcolor{green!20}1 &\cellcolor{green!40}2&\cellcolor{green!60}3&\cellcolor{green!80}max&\multicolumn{3}{c}{highest similarity}\\
		\end{tabular}
	}
	\caption{Similarity difference heat-map of xPPU scenarios: similarity is higher than with the coarse-grained metric marked green and lower marked red.}
	\label{tab:xppu_similarity_diff}
\end{table}

To investigate the impact of a fine-grained metric on the overall similarity, we created a heat-map that shows how the similarity values change between the comparison with fine-grained implementation artifacts with the fine-grained metric, compared to the coarse-grained metric that compares languages with count metrics.
In \autoref{tab:ppu_similarity_diff} and \autoref{tab:xppu_similarity_diff} we illustrate these maps for the PPU and xPPU scenarios.
Red-colored cells show that the fine-grained metric results in a lower similarity value than using the coarse-grained metric. A green-colored cell means that the resulting similarity is higher.  
As we can see, most of the cells are colored red, which means that the comparison utilizing the fine-grained comparison metric results in a lower similarity, e.g., the similarity between S17 and S19 is 3 percent lower.
We analyzed the resulting family models to explain and discuss this behavior.
For example, when we compare the scenarios S17 and S19, we can see that the \emph{Main} \ac{POU} implementation of S17 contains 93 statements and the scenario S19 95 statements.
With a simple statement count metric, which relates the statements, we obtained a similarity of $\frac{93}{95}\cdot 100 = 97,89\%$.
In contrast, the fine-grained comparison metric compares every single statement. 
In this case, we have 32 mandatory \mandatory{}, 61 altered \alternative{} statements, and 2 optional \optional{} statements.
Based on the changes, the resulting similarity between these implementations using a fine-grained metric is $68.72\%$.
Moreover, the comparison of nested languages can lower the similarity based on changed artifacts in the nested implementations. 
In the case that the nested artifacts are not altered, the overall similarity would not change.
Still, in the case that the nested artifacts are changed, it would yield a lower similarity.
Hence, we conclude that fine-grained metrics lower the overall similarity if fine-grained artifacts are altered.
When only adding or removing artifacts, the coarse-grained metric will capture the similarity correctly as well.

\paragraph{\textbf{Cloned POUs in the PPU and xPPU Scenarios (RQ 3.2)}}
A common approach to evolve systems is code cloning.
We analyzed each of the \ac{PPU} and \ac{xPPU} scenarios to identify possible cloned \acp{POU}.
To filter our results, we used a similarity threshold of 70\%, which means all artifacts that have a similarity lower than the threshold are not considered.
For each scenario, we performed a pairwise \ac{POU} comparison, which results in an analysis of 354 family models for the \ac{PPU} scenarios and 810 family models for the \ac{xPPU} scenarios.
\begin{figure}
	\centering
	\includegraphics[width=\linewidth]{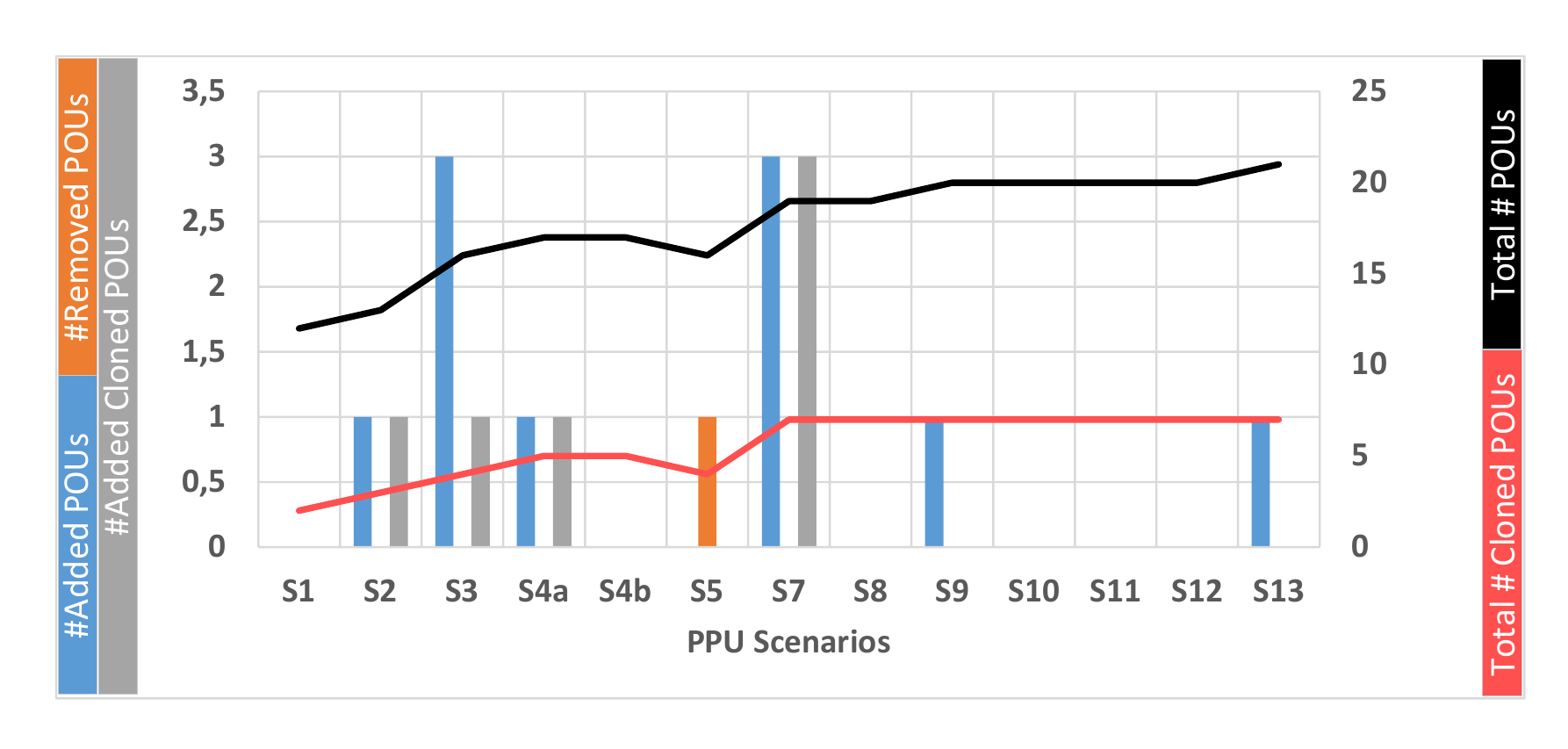}
	\caption{Results of the intra clone detection with all \ac{PPU} scenarios.}
	\label{fig:intra_results_ppu}
\end{figure}
\begin{figure}
	\centering
	\includegraphics[width=\linewidth]{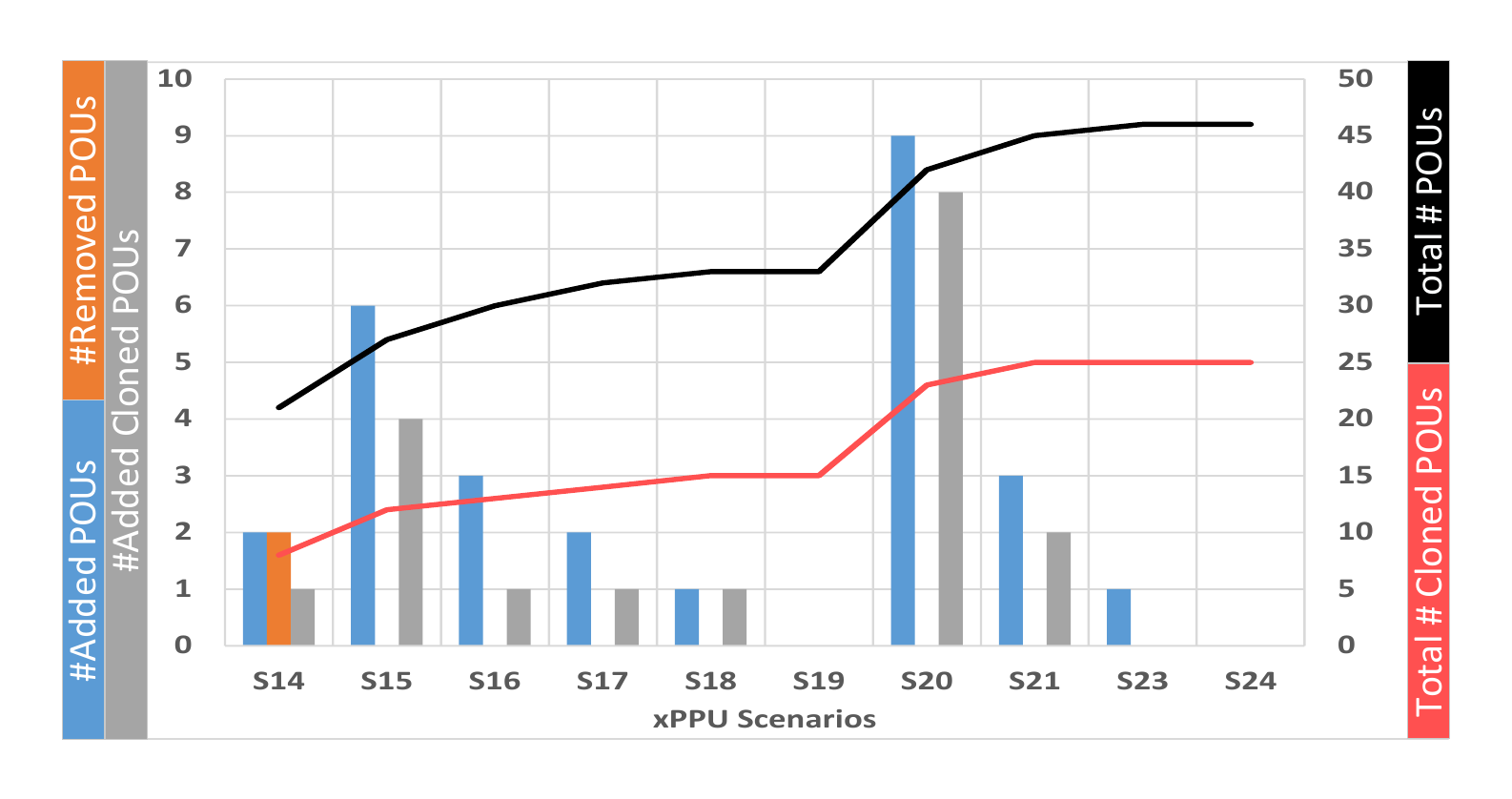}
	\caption{Results of the intra clone detection with all \ac{xPPU} scenarios.}
	\label{fig:intra_results_xppu}
\end{figure}
In \autoref{fig:intra_results_ppu} and \autoref{fig:intra_results_xppu} we show our aggregated results of the analysis of the \ac{PPU} and \ac{xPPU} scenarios.
The bars show the changed amount of \acp{POU} in the respective evolution step.
The blue bar shows how many \acp{POU} were added, the orange bar how many were removed, and the gray bar shows how many clones were introduced.
Besides, the black trend line shows the amount of \acp{POU} in the respective scenario, and the orange trend line shows how many of this \acp{POU} are potential clones.
The first scenario S1 contains a potential cloned \ac{POU} which stores the status of the respective \ac{POU} in a Boolean variable.
During the evolution of this system, nine \acp{POU} are added, in which 7 are potential clones.
For example, in scenario S3, the BistableCylinder is added, a Type~III clone of the MonostableCylinder.
\begin{figure}
	\centering
	\includegraphics[width=\linewidth]{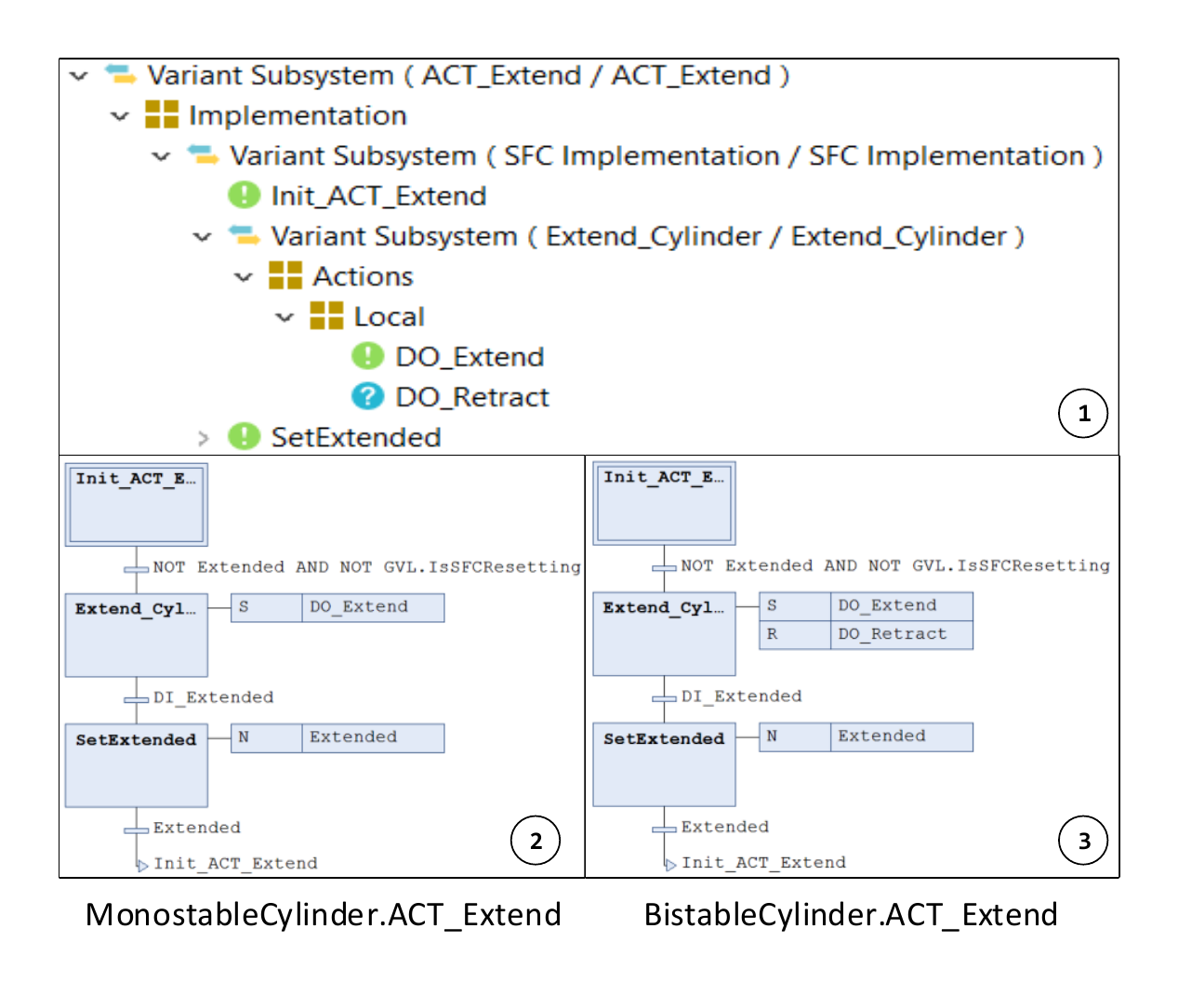}
	\caption{Resulting family model of the ACT\_Extend Action utilized by Monostable- and Bistablecylinder within scenario three.}
	\label{fig:scenario3_altered_action}
\end{figure}
In \autoref{fig:scenario3_altered_action} we depict a snippet of the resulting family model with the respective \ac{SFC} implementation of the comparison between the MonostableCylinder and BistableCylinder.
\circlearound{1} shows the family model of the ACT\_extend action used by both \acp{POU}.
As we can see, the \ac{SFC} implementation was altered by adding an action to the Extended\_Cylinder step. 
The difference can be inspected in the respective implementations shown in \circlearound{2} and \circlearound{3}.
An additional copy of the MonostableCylinder \ac{POU} is added during the evolution to scenario 15. 
The only change is a renaming of the \ac{POU} from Monostable Cylinder to Switch, which can be classified as a type II clone. 
In scenario 15, the \acp{POU} PicAlpha and RefillConveyor are added, which are clones of each other.
However, we were able to identify 25 of 46 \acp{POU} as potential clones in scenario 24, which represents the last evolution step.
Overall, we could detect Type~I and Type~III clones during the PPU and xPPU scenarios' evolution.

The intra- and inter-system clone detection approach allows visualization of cloned artifacts within a system and shows variation points between artifacts.
That information is useful for the developer to show potential candidates for the extraction of library components, which can be reused in different IEC~61131-3 systems.

\subsection{Threats to Validity}
This section categorizes the threats to the validity of our results into construct, internal, external, and conclusion validity and identifies the respective threats.

\subsubsection{Construct Validity}
To measure the precision, we used a mutation framework that generates mutant models.
Mutant models are syntactically correct, but we can not guarantee that they are semantically correct.
Measured precision and recall may be lower for a real-world subject system.

To identify the performance of our approach, we benchmarked the \ac{VAT}.
Therefore, we performed a run-time measure using the event-based benchmark system provided by the \ac{VAT}.
For memory consumption, we analyzed the memory dump that is created by the \emph{VisualVM}.
Both measures are key factors for the analysis of the scalability and are specially selected to give an insight into the VAT performance.
Other measures might also support the results in terms of performance, and other developers might identify more suitable ones.
Moreover, in a non-closed system, run-time deviations occur that we can not influence, such as hardware faults.
However, we argue that our selection of measures supports the corresponding performance analysis.
Furthermore, we measured the run-time several times and used the average value to reduce the deviation. 

To investigate our approach's usefulness, we manually evaluated the results of the \ac{VAT} using human intuition.
The resulting similarity values between element pairs and our matching approach's results may only make sense for us.
The intuition of other researchers or domain experts might be different, such that they question the usefulness of our results.
Besides, the metrics we used for comparing IEC~61131 programming languages and their nestings may not be suited to all scenarios.

\subsubsection{Internal Validity}
All measures we discussed and the data we gathered during the evaluation might be influenced by factors, which we did not consider.
This might restrict the validity of the performance measures and the drawn conclusions. 
However, we carefully analyzed the performance and performed control measures to reduce the random impact factors such as memory scheduling or the code optimization performed by the \ac{JVM}.

\subsubsection{External Validity}
We evaluated the comparison approach and especially the attributes by using synthetic case studies that show a representative history of changes.
Other scenarios can show different kinds of changes and nesting of languages, which we did not consider and may stay undetected, e.g., changing a statement type by replacing a while-loop with a for-loop.
We used the \ac{PPU} and \ac{xPPU} case studies to perform the comparative measurement.
This case study is only implemented in \ac{ST} and \ac{SFC}.
Systems of other domains may show other utilized programming languages and language nestings. 
We argue that evaluated scenarios show a significant complexity and are representative for systems in the domain of \acp{aPS}.

\subsubsection{Conclusion Validity}
Our comparison approach for IEC~61131-3 languages is realized using metrics to compare different variants of software systems.
Metrics use attributes, options, and weights and are highly configurable.
Also, the user can customize the \ac{VAT} by adjusting different thresholds in the preferences.
The configuration of the metric and setup of the \ac{VAT} largely depends on human intuition and the user's domain knowledge.
Consequently, the metrics we used during the evaluation are not reliable because other developers might select other parameters that can cause different results.
However, we argue that the used metrics and settings were created with high caution and are evaluated using a mutation framework.
Besides, we explicitly provide the capability to customize metrics and configure the \ac{VAT} to meet the user's expectations.

\section{Related Work} 	 			
\label{sec:related_work}			
Clone and own is a common and popular reuse strategy in the software development domain.
In the past decades, the interest in code clones is also exhibited in existing research's wealth.
In general, clone-detection aims to reduce large software systems' maintenance effort by tracing clones or transferring a software system into an \ac{SPL} \cite{juergens2009code}.
Both activities require a detailed analysis of the respective software systems.
Most of the research focused on detecting code clones in high-level programming languages such as C, C++, and Java \cite{mondal2020survey,roy2007survey,bellon2007comparison}.
However, in the domain of \acp{aPS}, cloning code is also a common practice \cite{DurdikSustainability,Legat:2013}.
To bridge the gap, we developed a comparison approach that allows detecting clones in and between software variants, which are implemented regarding the IEC61131-3 standard.
Our approach relates to two main categories, which are \emph{Intra Clone Detection} and \emph{Variability Analysis} or also referred to as \emph{Inter Clone Detection}.

\subsection*{Intra Variant Clone Detection (Classic Clone Detection)}
Different code clone detection approaches have been presented in the last decades.
Roy et al. classified clone detection techniques by utilizing internal source code representation, which are text-, token-, tree-, graph and model-based \cite{roy2007survey}. 
Kelter et al. \cite{kelter2005generic} show an approach to compare pairs of \ac{UML} class diagrams.
They semantically lift derived differences to enhance comprehensibility.
In extension of their work, they utilize \emph{State Charts} \cite{Kelter:2008:STATE}.
However, unlike our approach, results are not presented in a family model, in which we present our results.
Alanen et al.\cite{alanen2003difference} presented a model comparison of \ac{UML} models to achieve a model versioning system for MOF-based models.
They presented three meta-model independent algorithms to calculate differences, merge and calculate the union of two models, and depend on the UML's \acp{UUID}.
Results of the analysis are presented as deltas, which are used for the merge algorithm.

In the industrial domain, the analysis of behavior and data-flow models such as MATLAB/Simulink models that are often used in the context of \ac{MDE} becomes essential. 
Deissenboeck et al. \cite{deissenboeck2010model,deissenboeck2008clone} showed a clone detection approach for model-based languages on MATLAB/Simulink models using a depth-first search heuristic for cloned pairs on a labeled graph that is created out of a model.
Pham et al. \cite{pham2009complete} presented an accurate clone detection approach in graph-based models.
Those models are comparable with \ac{FBD} , which are using \emph{transitions} to connect \emph{blocks} to model a behavior.
However, both approaches transforming the models into a graph representation in the difference we transform the PLCOpenXML into a model and work directly on them.
Alalfi et al. \cite{alalfi2012near,alalfi2012models} adapted a text-based code clone detection technique to identify clones in the textual representation of Simulink models.
A particular issue they have is the graphical representation of results.
Our approach uses a solution structure, which can be presented as a family or technical feature model.

Yu et al. present an approach that detects clones in Java using the \emph{Smith-Waterman} algorithm on Java bytecode \cite{yu2019detecting}.
This approach is only suitable for languages that are translated into Java bytecode. 
In contrast to Java, IEC61131-3 languages are compiled into machine language that is executed directly and not interpreted by a \ac{JVM}.
Hummel et al. show an incremental graph-based clone detection algorithm in \cite{Hummel:2011}.
They use a data structure called the \emph{clone index} to store sub-graphs for efficient matching. 
Although the relation between statements forms a graph, adding or removing a statement changes the underlying graph. 
Consequently, the detection is inappropriate for variability mining, as we want to quantify possible clones in terms of similarities.
In \cite{kamiya2002ccfinder}, Kamiya et al. show a clone detection technique that uses transformation rules and a token-based comparison.
Their approach is limited and can not use source files written in two or more programming languages.
For our case, it is not suitable because IEC61131-3 source files can contain a mix of all four languages.
With the rise of deep learning, techniques were presented with a learning-based paradigm.
White et al. \cite{white2016deep} specified a learning-based approach that utilizes a \ac{RvNN} to detect clones in Java code.

In the domain of \ac{aPS} only a few researchers analyzed IEC 61131-3 software system.
H K et al. \cite{hk2019analysis} proposes a method for detecting semantic clones in IEC61131-3 based systems by performing an input-output variable impact and dependency analysis.
Thaller et al. \cite{thaller2017exploring} analyzed a real-world \ac{PLC} software system using an extended version of Simian \cite{harris2003simian}, which is a text-based clone detection tool.
The analyzed system is implemented using \ac{ST} and C++.
Fahimipirehgalin et al. \cite{fahimipirehgalin2019similarity} presented a call-graph based approach that allows 
detect similarities between two software structures, e.g., \acp{POU}, which support the detection of clones with a
pre-selection of potentially similar artifacts.
In the field of clone detection on \ac{LD} Nedvěd et al. \cite{nedvved2015tool} presents a tool that can detect differences between \ac{LD} implementation based on their PLCOpenXML representation.
We are the first to analyze IEC61131-3 variants with nested implementations on a fine-grained level to the best of our knowledge.

\subsection*{Inter Variant Clone Detection (Variability Analysis)}
The identification of variability in software evolution has been subject to research for several years. 
However, there are few extractive approaches focused on reverse engineering of legacy systems without feature information.
In our prior work, we analyze Pascal programs to identify code clones \cite{rosiak2019analyzing}.
We use a model-based approach for the comparison of Pascal variants that compares programs statement-wise.
For the comparison of \ac{ST} implementations, we adopt this approach because both languages have a similar structure.
Schlie et al. \cite{SCHLIE:SPLC17} employ an adjustable matching window technique to enhance variability analysis in \textit{MATLAB Simulink} models.
The windows define a sub-graph on which a data flow based comparison is applied to compare each model. 
The approach takes hierarchical structures of blocks into account while comparing windows.
In contrast to our approach, we strictly compare the four languages' implementation artifacts within the same hierarchical level.
Holthusen et al. \cite{holthusen2014family} showed a family model mining approach for \ac{FBD}, which is an adaption of their previous work in the automotive domain analyzing MATLAB Simulink models~\cite{wille2013interface,holthusen2013automatische}.
Both approaches are working on block-based languages, which have a more straightforward structure than textual languages.
Moreover, they do not consider the \ac{ST} implementation that can occur in some function blocks.
Duszynski et al. \cite{duszynski2011}, present the \textit{Variant Analysis}, which is an approach for the individual analysis of multiple software variants.
Variants are mapped onto a system structure model.
Although the variants are represented as models, they use a string-based comparison approach applied to these variants' source code.
As our model stores elements with their corresponding source code, we support string-based comparisons as well. 
However, our approach operates on pairs of software variants and compares them to a lower detail level.
Fischer et al. \cite{fischer2015ecco}, present the ECCO (Extraction and Composition for Clone-and-Own) tool, which can automatically locate reusable parts from previously developed variants.
The conceptual framework behind ECCO is shown in \cite{fischer2014enhancing}.
The process is divided into three steps: extraction, composition, and completion, while the variability analysis is performed on an artifact tree that allows comparing any artifacts.
The ECCO tool compares artifact trees and stops if two nodes are not equal.
In contrast to ECCO, our approach determines the changes and compares all sub-elements if it finds differences between artifacts.
Finally, our work's fundamental is the \ac{VAT}\cite{schlie2019analyzing} that we extended with our concepts and used to perform the evaluation.

\subsection*{Mutation Framework}
For the evaluation of or approach, we implemented a mutation framework that allows to mutate scenarios based on a meta-model and store the mutation context, which can be used as ground truth.
The mutation framework is inspired by other authors that also used a mutation based strategy to evaluate their clone-detection approach.
Roy et al. \cite{roy2009mutation} proposed an evaluation framework that uses code fragment mutation to create and inject known code clones that can be used to measure recall and precision of clone detection tools accurately. 
Stephan et al. \cite{stephan2012towards,stephan2013using} showed a mutation-based evaluation framework for evaluating the Simulink clone detection approach.
Svajlenko et al. \cite{svajlenko2015evaluating} introduced \emph{Big Clone Bench} as a big data, varied and comprehensive clone benchmark for modern tools.
In contrast to our work, the existing mutation frameworks use existing code fragments from a repository to inject them into models.
We use existing models as a seed and generate artifacts based on our meta-model, reflecting the IEC61131-3 standard.
Other authors use mutation frameworks in another context, such as software testing.
Just \cite{just2014major} showed Major, a framework for the mutation analysis and fault seeding to evaluate software test suits.

\section{Conclusion and Future Work}
\label{sec:conclusion}

With an increasing interest in variant variety for industrial products, variability has become a key factor of many software systems.
In the domain of \acp{aPS} and their control, software often remains in use for decades.
To reduce such a system's maintenance effort, the detection of clones and analysis of variability is crucial.
On the one hand, code-clones can be refactored into reusable artifacts such as library components.
And on the other hand, the variability analysis can support experts in migrating a system portfolio into an \ac{SPL}. 
To this end, the identification of code clones and the detection of variability in the domain of \acp{aPS} is a remaining challenge that requires appropriate tool support.
A key feature of such tools is a fine-grained analysis of implementation artifacts to provide useful results to domain experts.

This paper proposed a comparison approach for IEC~61131-3 languages to detect fine-grained changes between variants (Inter Clone Detection) and within a variant (Intra Clone Detection).
We implemented our comparison approach in the \ac{VAT} tool. 
To compare fine-grained implementation artifacts, we implemented 29 attributes and added options to compare nested implementation languages.
We evaluated our concept based on our implementation.

To assess our concept, we performed a qualitative and quantitative analysis.
The quantitative analysis allows us to reason about the correctness and performance of our approach.
Therefore, a mutation framework was implemented, which generates mutants that for automatic evaluation of the \ac{VAT}.
We applied a fine-grained and a coarse-grained metric and compared their impact on precision and recall.
To argue about the performance and scalability, we measured the run-time and memory consumption.

To show our clone-detection approach's usefulness as a step during the re-engineering of legacy \ac{aPS} systems,
we used two metrics during the qualitative analysis and applied them to all pairwise comparisons of the \ac{PPU} and \ac{xPPU} scenarios.
We showed the similarities for the \ac{PPU} and \ac{xPPU} scenarios and what impact the different granularities of metrics have on the similarity values and the precision and recall.

Moreover, we evaluated the detection and comparison of nested implementation languages.
In general, we conclude that our comparison approach with a fine-grained metric has excellent precision and recall values and can detect changed artifacts of nested implementations. 
Overall, the evaluation shows that our approach can detect fine-grained changes between and within IEC~61131-3 implementations down to fine-grained artifacts such as statements.

Our work provides several possibilities for future work.
To this point, we can only compare, match, and merge two variants.
For the generation of a 150\% model, it is crucial to extend our approach to cope with multiple input models.
This allows us to merge a complete variant portfolio and reduce the effort of migrating such systems.
Based on the clone detection results, we try to establish a semi-automatic refactoring of code-clones into reusable artifacts such as library components.
An additional field of interest is improving our mutation framework with more complex mutation operations.
More configuration possibilities such as only \ac{ST} or \ac{SFC} mutations would allow assessing the precision and recall of the clone detection for a single language.
This would help compare our approach with other solutions that only support the comparison of a subset of our approach.
Moreover, it eases the assessment of code detection tools and calculates precision and recall, but it can also be used to assess test suits.

\begin{acronym}
	\acro{aPS}{automated production system}
	\acroplural{aPS}{automated production systems}
	\acro{SPL}{software product line}
	\acroplural{SPLs}{software product lines}
	\acro{POU}{program organization unit}
	\acro{JVM}{Java Virtuel Machine}
	\acroplural{POUs}{program organization units}
	\acro{PLC}{programmable logic controller}
	\acro{ST}{Structured Text}
	\acro{FBD}{Function Block Diagram}
	\acro{LD}{Ladder Diagram}
	\acro{SFC}{Sequential Function Chart}
	\acro{VAT}{Variability Analysis Toolkit}
	\acro{PPU}{Pick and Place Unit}
	\acro{xPPU}{Extended Pick and Place Unit}
	\acro{IL}{Instruction List}
	\acro{MOF}{Meta-Object Facility\texttrademark}
	\acro{EMOF}{Essential Meta-Object Facility}
	\acro{OMG}{Object Managment Group}
	\acro{ANTLR}{ANother Tool for Language Recognition}
	\acro{RCP}{Rich Client Platform}
	\acro{EMF}{Eclipse Modeling Framwork}
	\acro{UML}{Unified Modeling Language }
	\acro{RvNN}{Recursive Neural Network}
	\acro{UUID}{Universally Unique Identifier}
	\acroplural{UUID}{Universally Unique Identifiers}
	\acro{MDE}{Model-driven engineering}
\end{acronym}

\bibliography{CCD_FOR_IEC_61131-3}

\end{document}